\documentclass[article]{aa}
\usepackage{graphicx}
\usepackage[figuresright]{rotating}
\usepackage{subfigure}
\usepackage{float}
\usepackage{txfonts}
\begin{document}
\title{The influence of chemical composition on the properties of 
Cepheid stars.}

   \subtitle{II -- The iron content\thanks{Based on observations made 
    with ESO Telescopes at Paranal and La Silla Observatories under 
    proposal ID 66.D-0571}}

   \author{M. Romaniello \inst{1}
          \and F. Primas \inst{1}
          \and M. Mottini \inst{1}
          \and S. Pedicelli\inst{1,2}
          \and B. Lemasle \inst{3} 
          \and G. Bono \inst{1,2,4}
          \and P. Fran\c{c}ois \inst{5}
          \and M.A.T. Groenewegen \inst{6}
          \and C. D. Laney\inst{7}
           }

   \institute{European Southern Observatory, Karl-Schwarzschild-Strasse 2, D-85748 Garching bei M{\"{u}}nchen, Germany
          \and Universit\`a of Roma Tor Vergata, Department of Physics, via della Ricerca Scientifica 1, I-00133 Rome, Italy
          \and Universit\'e de Picardie Jules Verne, Facult\'e des Sciences, 33 rue Saint-Leu, 80039 Amiens Cedex 1, France 
          \and INAF-Osservatorio Astronomico di Roma, via Frascati 33, I-00040 Monte Porzio Catone, Italy
          \and Observatoire de Paris-Meudon, GEPI, 61 avenue de l'Observatoire, F-75014 Paris, France
          \and Royal Observatory of Belgium, Ringlaan 3  B-1180 Brussels, Belgium 
          \and South African Astronomical Observatory, PO Box 9, 7935 Observatory, South Africa} 

   \date{Received ; accepted }

\abstract
%context heading (optional)
{The Cepheid period-luminosity (PL) relation is 
unquestionably one of the most powerful tools at our disposal for 
determining the extragalactic distance scale. While significant 
progress has been made in the past few years towards its understanding 
and characterization both on the observational and theoretical sides, 
the debate on the influence that chemical composition may have on the 
PL relation is still unsettled.} 
%aims heading (mandatory)
{With the aim to assess the influence 
of the stellar iron content on the PL relation in the $V$ and $K$ bands, 
we have related the $V$-band and the $K$-band residuals 
from the standard PL relations of Freedman et al. (2001) and Persson et 
al. (2004), respectively, to [Fe/H].} 
%method
{We used direct measurements of the 
iron abundances of 68 Galactic and Magellanic Cepheids from FEROS and 
UVES high-resolution and high signal-to-noise spectra.} 
%results heading (mandatory)
{We find a mean iron abundance ([Fe/H]) about solar ($\sigma$ = 0.10) for our 
Galactic sample (32 stars), $\sim$ -0.33 dex ($\sigma$ = 0.13) for the 
Large Magellanic Cloud (LMC) sample (22 stars) and $\sim$ -0.75 dex 
($\sigma$ = 0.08) for the Small Magellanic Cloud (SMC) sample (14 stars). 
Our abundance measurements of the Magellanic Cepheids double the number 
of stars studied up to now at high resolution. 
The metallicity affects the $V$-band Cepheid PL relation and metal-rich 
Cepheids appear to be systematically fainter than metal-poor ones. These 
findings depend neither on the adopted distance scale for Galactic Cepheids 
nor on the adopted LMC distance modulus.  
Current data do not allow us to reach a firm conclusion concerning the 
metallicity dependence of the $K$-band PL relation. The new Galactic distances 
indicate a small effect, whereas the old ones support a marginal effect.
}
%Conclusions
{Recent robust estimates of the LMC distance and current results  
indicate that the Cepheid PL relation is not Universal. 
}
{ \keywords{Cepheids -- Stars: abundances-- Stars: distances -- Stars: oscillations} } 

\titlerunning{The impact of chemical composition on Cepheid properties}

\authorrunning{Romaniello et al.}

\offprints{G. Bono, e-mail: {\tt bono@mporzio.astro.it}}

   \maketitle
%
%________________________________________________________________

\section{Introduction}
Since the dawn of modern astronomy the Cepheid Period-Luminosity 
(PL) relation is a key tool in determining Galactic and extragalactic 
distances. In spite of its fundamental importance, the debate on the 
role played by the chemical composition on the pulsation properties 
of Cepheids is still open, with different theoretical models and observational 
results leading to markedly different conclusions. 

From the theoretical point of 
view pulsation models by different groups lead to substantially different 
results. Linear models (e.g. Chiosi et al. 1992; Sandage et al. 1999; Baraffe
\& Alibert 2001), based on nonadiabatic radiative models, suggest a moderate 
dependence of the PL relation on the metallicity. The predicted change at 
log(P) = 1 is less than 0.1 mag for metal abundances ranging from the SMC 
(Z=0.004) to the Galaxy (Z =0.02), independent of wavelength. Nonlinear 
convective models (e.g. Bono et al. 1999; Caputo et al. 2000; Caputo 2008) 
instead predict a larger dependence on the same interval of metallicity: the 
change is 0.4 mag in $V$, 0.3 mag in $I$ and 0.2 mag in $K$, again at log(P) = 1. 
Moreover, the predicted change in these latter models is such that metal-rich 
Cepheids are fainter than metal-poor ones, at variance with the results of the 
linear models. Fiorentino et al. (2002) and, more recently, Marconi, Musella \& 
Fiorentino (2005) investigations, also based on nonlinear models, suggest that 
there may be also a dependence on the helium abundance.

On the observational side, the majority of the constraints comes from 
indirect measurements of the metallicity, mostly in external galaxies, 
such as oxygen nebular abundances derived from spectra of H II regions 
at the same Galactocentric distance as the Cepheid fields  
(e.g. Sasselov et al. 1997; Kennicutt et al. 1998; Sakai et al. 2004). 
These analyses indicate that metal-rich Cepheids are brighter than 
metal-poor ones (hence at variance with the predictions of nonlinear 
convective models), but it is important to note that the results span 
a disappointingly large range of values (see Table~1 and Fig.~1). 

More recently Macri et al. (2006) found, by adopting a large sample 
of Cepheids in two different fields of NGC~4258 and the [O/H] gradient 
based on H II regions provided by Zaritsky et al. (1994), a metallicity 
effect of $\gamma=-0.29\pm0.09$ mag/dex. This galaxy has been adopted 
as a benchmark for estimating the metallicity effect, since an accurate 
geometrical distance based on water maser emission is available 
(Herrnstein et al. 2005). However, in a thorough investigation Tammann 
et al. (2007) suggested that the flat slope of the Period-Color relation 
of the Cepheids located in the inner metal-rich field could be due to a 
second parameter, likely helium, other than the metal abundance.  
Furthermore, Bono et al. (2008) found, using the new and more accurate nebular 
oxygen abundances for a good sample of H II region in NGC~4258 provided 
by D\`iaz et al. (2000), a shallower abundance gradient. In particular, the 
new estimates seem to suggest that both the inner and the outer field 
might have a mean oxygen abundance very similar to LMC. They also found 
a very good agreement between predicted and observed Period-Wesenheit 
($V,I$) relation. Nonlinear convective models predict for this relation 
a metallicity effect of $\gamma=+0.05\pm0.03$ mag/dex.   

On the basis of independent distances for 18 galaxies based on Cepheid 
and on the Tip of the Red Giant Branch, Tammann et al. (2007) found a 
small metallicity effect ($\gamma=-0.017\pm0.113$ mag/dex). On the 
other hand, Fouqu\'e et al. (2007) using a sample of 59 Galactic 
Cepheids whose distances were estimated using different methods
-- HST trigonometric parallaxes (Benedict et al. 2007), revised 
Hipparcos parallaxes (van Leeuwen et al. 2007), infrared surface 
brightness method (Fouqu\'e \& Gieren 1997), and interferometric 
Baade-Wesselink method (Kervella et al. 2004), zero-age-main-sequence
fitting of open clusters (Turner \& Burke 2002) -- found no significant 
difference between optical and Near-Infrared (NIR) slopes of Galactic and LMC 
Cepheids (Udalski et al. 1999; Persson et al. 2004).

\begin{figure}[h]
\centering
\includegraphics[height=0.35\textheight, width=0.40\textwidth]{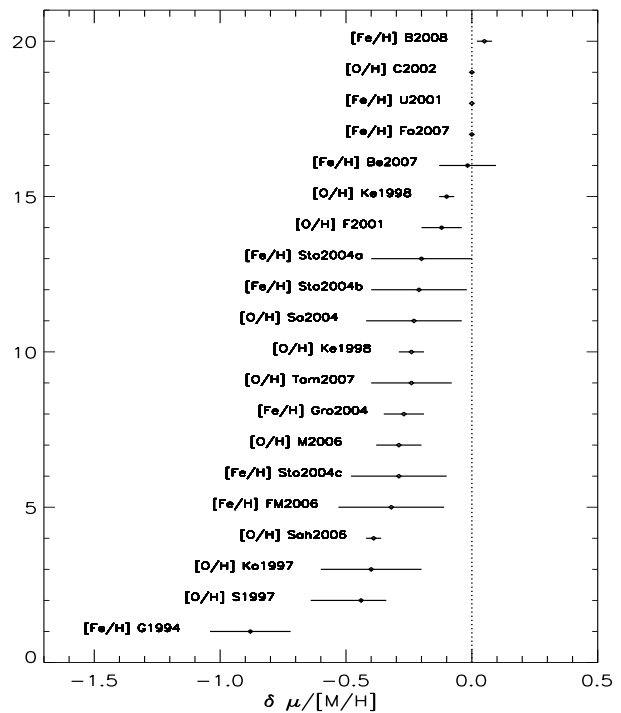} \vspace*{0.5truecm} 
\caption{
Comparison of recent results for the metallicity sensitivity 
of Cepheid distances. FM1990: Freedman \& Madore (1990); 
G1994: Gould (1994); Ko1997: Kochanek (1997); S1997: 
Sasselov et al (1997); Ke1998: Kennicutt et al (1998); F2001: 
Freedman et al (2001); U2001: Udalski et al (2001); C2002: 
Ciardullo et al (2002);  Sa2004: Sakai et al (2004); Sto2004: 
Storm et al (2004); Gro2004: Groenewegen et al (2004); 
M2006: Macri et al. (2006); Sah2006: Saha et al. (2006); 
Be2007: Benedict et al. (2007); Fo2007: Fouqu\'e et al. (2007); 
Tam2007: Tammann et al. (2007); B2008: Bono et al. (2008). 
See Table~1.
}
\label{Fig1}
\end{figure}

An alternative approach is to measure directly 
the metal content of Cepheid stars, which, so far, has been attempted 
only by few studies, primarily focused on stars of our own Galaxy (Luck 
\& Lambert, 1992; Fry \& Carney 1997; Luck et al. 1998; Andrievsky et al. 
2002a,b,c; Luck et al. 2003; Andrievsky et al. 2004). Fry \& Carney (1997, 
hereafter FC97), for instance, have derived iron and $\alpha$-element abundances 
for 23 Galactic Cepheids from high resolution and high signal-to-noise spectra. 
They found a spread in [Fe/H] of about 0.4 dex, which they claim is real. 
Using approximately half of their sample, the stars belonging to clusters 
or associations, they have made a preliminary evaluation of metallicity effects 
on the zero point of the PL relation, finding that metal-rich Cepheids are brighter 
than metal-poor ones. Thus, finding a result similar to the studies based on 
indirect measurements of the metallicity.

The impressive observational effort carried out by Andrievsky and collaborators 
(Andrievsky et al. 2002a, 2002b, 2002c; Luck et al. 2003; Andrievsky et al. 
2004; Kovtyukhet al. 2005b) has, instead, taken advantage of high resolution 
spectra of 130 Galactic Cepheids (collected with different instruments at 
different telescopes) in order to determine their chemical composition and 
study the Galactic abundance gradient. The sample covers a range of Galactocentric 
distances from 4 to 14 kpc. The emerging picture can be best described by a 
relatively steep gradient (about -0.14 dex kpc$^{-1}$) for Galactocentric 
distances less than 7 kpc, followed by a much shallower slope 
($\approx$ -0.03 dex kpc$^{-1}$) between 
7 and 10 kpc, a discontinuity at approximately 10 kpc and a nearly constant 
metallicity of about -0.2 dex towards larger Galactocentric distances, out 
to about 14 kpc. In relation to our work, it is important to note that Andrievsky 
and collaborators did not investigate the effects of the chemical composition on 
the Cepheid PL relation: on the contrary, they used the PL relation to determine 
the distances of their stars.

Outside our Galaxy, Luck \& Lambert (1992, hereafter LL92) have studied 10 
Cepheids in the Magellanic Clouds (MCs). Five are in the Large MC (LMC) 
and five in the Small MC (SMC). For the former sample, they found a mean 
[Fe/H] of -0.36 dex with a dispersion of 0.3 dex, while for the latter one 
the mean [Fe/H] is -0.60 dex with a rather small dispersion of less than 0.15 dex. 
A more recent study by Luck et al. (1998, hereafter L98) on 10 LMC Cepheids 
and 6 SMC Cepheids, 4 of which in common with LL92, confirmed the mean [Fe/H] 
value in the LMC (-0.30 dex), found very little evidence of a significant 
metallicity dispersion in the LMC (contrary to LL92, but similarly to the SMC), 
and slightly revised downwards the mean [Fe/H] of the SMC (-0.74 vs -0.60 
found by LL92).

\begin{table*}[!ht]
%\begin{sidewaystable*}
%\begin{minipage}[t][180mm]{\textwidth}
\label{table:1}
\caption{Overview of recent results for the metallicity sensitivity of Cepheid 
distances. In the first column is listed the variation of the distance modulus
$\mu$ per dex of metallicity, the negative sign indicates that the true distance 
is longer than the one obtained neglecting the effect of the metallicity. 
In the second column is listed the elemental abundance used as reference 
for the metallicity. The third and fourth columns give the method and the 
reference of the different studies. See also Fig.~1.}
%\centering
\begin{tabular}{ l c p{9cm} l} \hline
$\delta \mu / \delta [M/H]$ & ~ & Method & Reference \\ 
(mag/dex) & ~ & ~ \\ \hline\hline
-0.32 $\pm$ 0.21 & [Fe/H] & Analysis of Cepheids in 3 fields of M31 ($BVRI$ bands) & Freedman \& Madore (1990) \\
-0.88 $\pm$ 0.16 & [Fe/H] & Comparison of Cepheids from 3 fields of M31 and LMC ($BVRI$ bands) & Gould (1994) \\
-0.40 $\pm$ 0.20 & [O/H] & Simultaneous solution for distances to 17 galaxies ($UBVRIJHK$ bands) & Kochanek (1997) \\
-0.44$_{-0.20}^{+0.10}$ & [O/H] & Comparison of EROS observations of SMC and LMC Cepheids ($VR$ bands) & Sasselov et al. (1997) \\
-0.24 $\pm$ 0.16 & [O/H] & Comparison of HST observations of inner and outer fields of M101 & Kennicutt et al. (1998) \\
-0.12 $\pm$ 0.08 & [O/H] & Comparison of 10  Cepheid galaxies with Tip of the Red Giant Branch distances & Kennicutt et al. (1998) \\
-0.20 $\pm$ 0.20 & [O/H] & Value adopted for the HST Key Project final result & Freedman et al. (2001) \\
 0 & [Fe/H] & OGLE result comparing Cepheids in IC1613 and MC ($VI$ bands) & Udalski et al. (2001) \\
 0 & [O/H] & Comparison of Planetary Nebula luminosity function distance scale and 
          Surface Brightness fluctuation distance scale & Ciardullo et al. (2002) \\
-0.24 $\pm$ 0.05 & [O/H] &  Comparison of 17 Cepheid galaxies with Tip of the Red Giant Branch distances & Sakai et al. (2004) \\
-0.21 $\pm$ 0.19 & [Fe/H] &  Baade-Wesselink analysis of Galactic and SMC Cepheids ($VK$ bands) & Storm et al. (2004) \\
-0.23 $\pm$ 0.19 & [Fe/H] &  Baade-Wesselink analysis of Galactic and SMC Cepheids ($I$ band) &  Storm et al. (2004) \\
-0.29 $\pm$ 0.19 & [Fe/H] &  Baade-Wesselink analysis of Galactic and SMC Cepheids ($W$ index) &  Storm et al. (2004) \\
-0.27 $\pm$ 0.08 & [Fe/H] &  Compilation from the literature of distances and 
          metallicities of 53 Galactic and MC Cepheids ($VIWK$ bands) & Groenewegen et al. (2004) \\ 
-0.39 $\pm$ 0.03 & [Fe/H] &  Cepheid distances to SNe Ia host galaxies & Saha et al. (2006)\\ 
-0.29 $\pm$ 0.09 & [O/H]  & Cepheids in NGC~4258 and [O/H] gradient from Zaritsky et al. (1994)  & Macri et al. (2006)\\
-0.10 $\pm$ 0.03 & [Fe/H] & Weighted mean of Kennicutt, Macri and Groenewegen estimates & Benedict et al. (2007)\\
-0.017 $\pm$ 0.113 & [O/H] & Comparison between Cepheid and TRGB distances for 18 galaxies & Tammann et al. (2007)\\
 0  & [Fe/H] & Comparison between the slopes of Galactic and LMC Cepheids  & Fouqu\'e et al. (2007)\\
+0.05 $\pm$ 0.03  & [Fe/H] & Predicted Period-Wesenheit ($V,I$) relation & Bono et al. (2008)\\ \hline
\hline
\end{tabular}
%\vfill
%\end{minipage}
%\end{sidewaystable*}
\end{table*}

Finally, we mention two studies that followed slightly different 
approaches. Groenewegen et al. (2004) have selected from the literature a 
sample of 37 Galactic, 10 LMC and 6 SMC Cepheids for which individual 
metallicity estimates and $BVIK$ photometry were known. Their work aimed 
at investigating the metallicity dependence of the PL relation using 
individual metallicity determinations as well as good individual distance 
estimates for Galactic Cepheids. They inferred a metallicity effect of about 
-0.27 $\pm$ 0.08 mag/dex in the zero point in $VIWK$, in the sense that metal-rich 
Cepheids are brighter than the metal-poor ones (see Table~1 and Fig.~1, for a 
comparison with other studies). Also Storm et al. (2004) discussed the effect 
of the metallicity on the PL relation using 34 Galactic and 5 SMC Cepheids, 
for which they determined accurate individual distances with the Baade-Wesselink 
method. Assuming an average abundance for the SMC Cepheids of [Fe/H]= -0.7 and 
solar metallicity for the Galactic ones, they determined, in a purely differential 
way, the following corrections: -0.21 $\pm$ 0.19 for the $V$ and $K$ bands, -0.23 $\pm$ 0.19 
for the $I$-band and -0.29 $\pm$ 0.19 for the Wesenheit index W. These agree well 
with Groenewegen et al. (2004).

Despite these ongoing observational efforts, it is important to underline that none 
of the observational studies undertaken so far has directly determined elemental 
abundances of a large sample of Cepheids in order to explicitly infer the metallicity 
effect on the PL relation, taking advantage of a sample that has been homogeneously 
analysed.

The novelty of our approach consists exactly in this, i.e. in the homogeneous 
analysis of a large sample of stars (68) in
three galaxies (the Milky Way, the LMC and the SMC) spanning a factor of ten 
in metallicity and for which distances and $BVJK$ photometry are available. 
Preliminary results based on a sub-sample of the data discussed here were 
presented by Romaniello et al. (2005). Here we present the results about 
the iron content for the complete sample. In a forthcoming paper we will 
discuss the $\alpha$-elements abundances.

The paper is organised as follows. The data sample is presented in Section 2. 
In Section 3 we thoroughly describe the data analysis and how we determine the 
metallicity of our stars. We compare our iron abundances with previous results 
in Section 4. The dependence of the PL relation on [Fe/H] is discussed in Section 5. 
Finally, Section 6 summarizes our concluding remarks.

\begin{table*}
\label{table:2}
\caption{Pulsation phases ($\phi$) and intrinsic parameters of the Galactic 
Cepheids. Both AP Pup and AX Vel were not included in the analysis of the 
metallicity effect because accurate distance estimates are not available in 
literature. In particular, for AP Pup we only listed the apparent mean 
magnitudes. In the last column is listed the duplicity status according to 
Szabados (2003): B - spectroscopic binary, Bc - spectroscopic binary 
that needs confirmation, O - spectroscopic binary with known orbit, 
V - visual binary}
\centering 
\renewcommand{\footnoterule}{}
\begin{tabular}{l l c c c c c c c c c} \hline\hline
ID & $\log P$ & $\phi$ & $\mu_{Old}$  & $E(B-V)_{Old}$ & $\mu_{New}$  &  $E(B-V)_{New}$ & $M_{B}$ & $M_{V}$ & $M_{K}$ &  Duplicity \\ \hline

  l  Car  &     1.5509 & 0.580  &    8.99$^{d}$ &   0.170$^{d}$   &  8.56$^{b}$ & 0.147$^{a}$ &   -4.17	&    -5.28 	 &  -7.53    &  \ldots \\
  U  Car  &     1.5891 & 0.490  &   10.97$^{d}$ &   0.283$^{d}$   & 10.87$^{a}$ & 0.265$^{a}$ &   -4.50	&    -5.41 	 &  -7.44    &  B \\
  V  Car  &     0.8259 & 0.375  &    9.84$^{e}$ &   0.174$^{h}$   & 10.09$^{a}$ & 0.169$^{a}$ &   -2.54	&    -3.24 	 &  -4.86    &  B \\ 
 WZ  Car  &     1.3620 & 0.745  &   12.92$^{d}$ &   0.384$^{d}$   & 12.69$^{a}$ & 0.370$^{a}$ &   -3.80	&    -4.58 	 &  -6.52    &  \ldots \\
  V  Cen  &     0.7399 & 0.155  &    9.18$^{d}$ &   0.289$^{c}$   &  8.91$^{a}$ & 0.292$^{a}$ &   -2.41	&    -2.99 	 &  -4.49    &  \ldots \\
 KN  Cen  &     1.5319 & 0.867  &   13.12$^{d}$ &   0.926$^{d}$   & 12.84$^{a}$ & 0.797$^{a}$ &   -4.63	&    -5.46 	 &  -7.59    &  B \\  
 VW  Cen  &     1.1771 & 0.967  &   12.80$^{d}$ &   0.448$^{d}$   & 12.76$^{a}$ & 0.428$^{a}$ &   -2.93	&    -3.85 	 &  -6.08    &  B \\  
 XX  Cen  &     1.0395 & 0.338  &   11.11$^{d}$ &   0.260$^{d}$   & 10.90$^{a}$ & 0.266$^{a}$ &   -3.19	&    -3.91 	 &  -5.58    &  B \\
$\beta$ Dor  &  0.9931 & 0.529  &    7.52$^{c}$ &   0.040$^{c}$   &  7.50$^{b}$ & 0.052$^{a}$ &   -3.16	&    -3.91 	 &  -5.57    &  \ldots \\	
$\zeta$ Gem  &  1.0065 & 0.460  &    7.78$^{c}$ &   0.010$^{c}$   &  7.81$^{b}$ & 0.014$^{a}$ &   -3.16	&    -3.94 	 &  -5.72    &  V \\	 
 GH  Lup  &     0.9675 & 0.031  &   10.05$^{e}$ &   0.364$^{h}$   & 10.25$^{a}$ & 0.335$^{a}$ &   -2.77	&    -3.66 	 &  -5.54    &  B \\ 
  T  Mon  &     1.4319 & 0.574  &   10.82$^{d}$ &   0.209$^{c}$   & 10.71$^{a}$ & 0.181$^{a}$ &   -4.16	&    -5.15 	 &  -7.25    &  O \\
  S  Mus  &     0.9850 & 0.266  &    9.81$^{e}$ &   0.147$^{h}$   &  9.57$^{a}$ & 0.212$^{a}$ &   -3.48	&    -4.10 	 &  -5.62    &  O \\ 
 UU  Mus  &     1.0658 & 0.865  &   12.59$^{d}$ &   0.413$^{d}$   & 12.41$^{a}$ & 0.399$^{a}$ &   -3.12	&    -3.86 	 &  -5.70    &  \ldots \\
  S  Nor  &     0.9892 & 0.343  &    9.91$^{d}$ &   0.189$^{c}$   &  9.87$^{a}$ & 0.179$^{a}$ &   -3.23	&    -4.00 	 &  -5.77    &  B \\
  U  Nor  &     1.1019 & 0.422  &   10.72$^{d}$ &   0.892$^{d}$   & 10.46$^{a}$ & 0.862$^{a}$ &   -3.14	&    -3.90 	 &  -5.72    &  \ldots \\
  X  Pup  &     1.4143 & 0.232  &   12.36$^{e}$ &   0.443$^{h}$   & 11.64$^{a}$ & 0.402$^{a}$ &   -3.57	&    -4.38 	 &  -6.34    &  \ldots \\ 
 AP  Pup  &     0.7062 & 0.109  &  \ldots$^{f}$ &   \ldots$^{f}$  &\ldots$^{f}$ &\ldots$^{f}$ &    7.37	&     6.78	 &   5.26    &  B \\
 AQ  Pup  &     1.4786 & 0.436  &   12.52$^{d}$ &   0.512$^{c}$   & 12.41$^{a}$ & 0.518$^{a}$ &   -4.53	&    -5.35 	 &  -7.27    &  B \\
 BN  Pup  &     1.1359 & 0.397  &   12.95$^{d}$ &   0.438$^{c}$   & 12.93$^{a}$ & 0.416$^{a}$ &   -3.55	&    -4.33 	 &  -6.14    &  \ldots \\
 LS  Pup  &     1.1506 & 0.012  &   13.55$^{d}$ &   0.478$^{d}$   & 13.39$^{a}$ & 0.461$^{a}$ &   -3.60	&    -4.37 	 &  -6.18    &  B \\	   
 RS  Pup  &     1.6174 & 0.944  &   11.56$^{d}$ &   0.446$^{c}$   & 11.30$^{a}$ & 0.457$^{a}$ &   -4.71	&    -5.69 	 &  -7.81    &  \ldots \\
 VZ  Pup  &     1.3649 & 0.816  &   13.08$^{d}$ &   0.471$^{c}$   & 12.84$^{a}$ & 0.459$^{a}$ &   -3.93	&    -4.63 	 &  -6.31    &  \ldots \\
 KQ  Sco  &     1.4577 & 0.446  &   12.36$^{c}$ &   0.839$^{c}$   & 12.23$^{g}$ & 0.869$^{a}$ &   -4.05	&    -5.11 	 &  -7.55    &  \ldots \\
 EU  Tau  &     0.3227 & 0.414  &   10.27$^{c}$ &   0.170$^{c}$   & 10.27$^{c}$ & 0.170$^{d}$ &   -2.26	&    -2.74 	 &  -4.05    &  Bc\\
 SZ  Tau  &     0.4981 & 0.744  &    8.73$^{c}$ &   0.290$^{c}$   &  8.55$^{a}$ & 0.295$^{a}$ &   -2.38	&    -2.93 	 &  -4.33    &  B \\
  T  Vel  &     0.6665 & 0.233  &    9.80$^{d}$ &   0.281$^{c}$   & 10.02$^{a}$ & 0.289$^{a}$ &   -2.24	&    -2.88 	 &  -4.47    &  B \\
 AX  Vel  &     0.5650 & 0.872  &   10.76$^{f}$ &   0.224$^{h}$   &\ldots$^{f}$ & 0.224$^{a}$ &   \ldots&    \ldots 	 &  \ldots   &  \ldots \\
 RY  Vel  &     1.4496 & 0.704  &   12.02$^{d}$ &   0.562$^{c}$   & 11.73$^{a}$ & 0.547$^{a}$ &   -4.23	&    -5.05 	 &  -6.96    &  \ldots \\  
 RZ  Vel  &     1.3096 & 0.793  &   11.02$^{d}$ &   0.335$^{c}$   & 10.77$^{a}$ & 0.299$^{a}$ &   -3.78	&    -4.61 	 &  -6.56    &  \ldots \\
 SW  Vel  &     1.3700 & 0.792  &   11.00$^{d}$ &   0.349$^{c}$   & 11.88$^{a}$ & 0.344$^{a}$ &   -4.02	&    -4.83 	 &  -6.75    &  \ldots \\
 SX  Vel  &     0.9800 & 0.497  &   11.44$^{e}$ &   0.250$^{h}$   & 11.41$^{g}$ & 0.263$^{a}$ &   -3.33	&    -3.95 	 &  -5.49    &  \ldots \\ \hline  
\hline
\multicolumn{8}{l}{$^a$ Fouqu\'e et al. (2007).}\\
\multicolumn{8}{l}{$^b$ Benedict et al. (2007).}\\	   
\multicolumn{8}{l}{$^c$ Groenewegen et al. (2004).}\\    
\multicolumn{8}{l}{$^d$ Storm et al. (2004).}\\
\multicolumn{8}{l}{$^e$ Laney \& Stobie (1995) and Groenewegen (2004).}\\
\multicolumn{8}{l}{$^f$ Not included in the analysis of the metallicity effect.}\\
\multicolumn{8}{l}{$^g$ Groenewegen (2008).}\\
\multicolumn{8}{l}{$^h$ Fernie et al. (1995).}\\
\end{tabular}
\end{table*}

\begin{table*}
\label{table:3}
\caption{Pulsation phases ($\phi$) and intrinsic parameters of the Magellanic 
Cepheids. Periods ($\log P$), apparent mean magnitudes and reddenings come from 
Laney \& Stobie (1994). The mean $K$-band magnitudes were transformed into the  
2MASS photometric system using the transformation provided by Koen et al. (2007).}
\centering
\begin{tabular}{l r r r r r c} \hline\hline
ID & $\log P$ & $\phi$ & $B_0$ & $V_0$ & $K_0$ & $E(B-V)$\\ \hline
\multicolumn{7}{c}{LMC} \\ \hline
HV 877   &  1.654 & 0.682 & 14.06 & 12.98 &  10.77  &  0.12 \\
HV 879   &  1.566 & 0.256 & 14.12 & 13.15 &  11.03  &  0.06 \\
HV 971   &  0.968 & 0.237 & 14.86 & 14.24 &  12.68  &  0.06 \\
HV 997   &  1.119 & 0.130 & 14.94 & 14.19 &  12.37  &  0.10 \\
HV 1013  &  1.382 & 0.710 & 14.39 & 13.46 &  11.41  &  0.11 \\
HV 1023  &  1.425 & 0.144 & 14.48 & 13.51 &  11.45  &  0.07 \\
HV 2260  &  1.112 & 0.144 & 15.19 & 14.43 &  12.67  &  0.13 \\
HV 2294  &  1.563 & 0.605 & 13.19 & 12.45 &  10.74  &  0.07 \\
HV 2337  &  0.837 & 0.861 &\ldots & \ldots&  13.27  &  0.07 \\ 
HV 2352  &  1.134 & 0.201 & 14.49 & 13.84 &  12.25  &  0.10 \\
HV 2369  &  1.684 & 0.136 & 13.15 & 12.29 &  10.38  &  0.10 \\
HV 2405  &  0.840 & 0.037 & \ldots& \ldots&  13.43  &  0.07 \\
HV 2580  &  1.228 & 0.119 & 14.33 & 13.67 &  11.92  &  0.09 \\
HV 2733  &  0.941 & 0.411 & 14.85 & 14.34 &  13.00  &  0.11 \\
HV 2793  &  1.283 & 0.917 & 14.49 & 13.58 &  11.75  &  0.10 \\
HV 2827  &  1.897 & 0.880 & 13.19 & 12.03 &   9.80  &  0.08 \\
HV 2836  &  1.244 & 0.059 & 14.85 & 14.02 &  12.04  &  0.18 \\
HV 2864  &  1.041 & 0.055 & 15.16 & 14.42 &  12.77  &  0.07 \\
HV 5497  &  1.997 & 0.321 & 12.73 & 11.63 &   9.43  &  0.10 \\
HV 6093  &  0.680 & 0.024 & 15.74 & 15.16 &  13.71  &  0.06 \\
HV 12452 &  0.941 & 0.860 & 15.25 & 14.60 &  12.83  &  0.06 \\
HV 12700 &  0.911 & 0.342 & 15.62 & 14.87 &  13.12  & -0.01 \\ \hline
\multicolumn{7}{c}{SMC} \\ \hline
HV 817   &  1.277 & 0.298 &   14.13  & 13.59  &  12.12  &  0.08 \\
HV 823   &  1.504 & 0.873 &   14.46  & 13.60  &  11.58  &  0.05 \\
HV 824   &  1.818 & 0.315 &   13.06  & 12.27  &  10.33  &  0.03 \\ 
HV 829   &  1.931 & 0.348 &   12.61  & 11.81  &   9.92  &  0.03 \\ 
HV 834   &  1.866 & 0.557 &   12.95  & 12.14  &  10.20  &  0.02 \\
HV 837   &  1.631 & 0.822 &   13.95  & 13.10  &  11.11  &  0.04 \\
HV 847   &  1.433 & 0.500 &   14.40  & 13.66  &  11.83  &  0.08 \\
HV 865   &  1.523 & 0.108 &   13.55  & 12.93  &  11.21  &  0.06 \\
HV 1365  &  1.094 & 0.184 &   15.39  & 14.79  &  13.20  &  0.07 \\
HV 1954  &  1.223 & 0.847 &   14.13  & 13.62  &  12.12  &  0.07 \\
HV 2064  &  1.527 & 0.279 &   14.28  & 13.50  &  11.61  &  0.07 \\
HV 2195  &  1.621 & 0.135 &   13.85  & 13.07  &  11.09  & -0.02 \\
HV 2209  &  1.355 & 0.822 &   13.99  & 13.42  &  11.84  &  0.04 \\
HV 11211 &  1.330 & 0.516 &   14.36  & 13.64  &  11.83  &  0.06 \\ \hline
\hline
\end{tabular}
\end{table*}

\section{The data sample}

We have observed a total of 68 Galactic and Magellanic Cepheid stars.  
The spectra of the 32 Galactic stars were collected at the ESO 1.5m telescope 
on Cerro La Silla with the Fibrefed Extended Range Optical Spectrograph (FEROS, 
Pritchard 2004). Two fibres simultaneously feed the spectrograph: one fibre 
records the object spectrum, while the second one is fed either by the sky 
background or by a calibration lamp to monitor the instrument stability during 
the exposure (the sky background was chosen in our case). The CCD is a thinned, 
back illuminated detector with 15 micron pixels (2048 x 4096 pixels). 
The resolving power is 48,000 and the accessible wavelength range is from 
3,700 to 9,200 $\AA$.

The spectra of the 22 Cepheids in the LMC and the 14 Cepheids in the SMC were 
obtained at the VLT-Kueyen telescope on Cerro Paranal with the UV-Visual 
Echelle Spectrograph (UVES, Dekker et al. 2000; Kaufer et al. 2004) in service 
mode. The detector in the Red Arm is a mosaic of two CCDs (EEV + MIT/LL) with 
15 micron pixels (2048 x 4096 pixels). The resolving power is about 30,000 
(corresponding to a slit of 100) and the spectral range covered in our spectra 
is from 4800 to 6800 $\AA$ (580 nm setting).

The 2-D raw spectra were run through the respective instrument Data 
Reduction softwares, yielding 1-D extracted, wavelength calibrated and 
rectified spectra. The normalization of the continuum was refined with the 
IRAF task continuum. The 1- D spectra were corrected for heliocentric velocity 
using the rvcorr and dopcor IRAF tasks. The latter was also used to apply the 
radial velocity correction, which was derived from 20 unblended narrow lines 
well spread over the spectrum, selected among the species \ion{Fe}{i}, 
\ion{Fe}{ii} and \ion{Mg}{i}. The measured signal-to-noise ratios vary between 
70 and 100 for the FEROS spectra and between 50 and 70 for the UVES spectra.

The selected Cepheids span a wide period range, from 3 to 99 days. 
For the Galactic Cepheids we have adopted periods, optical and NIR photometry 
from Laney \& Stobie (1994), Storm et al. (2004), Groenewegen et al. (2004), 
Benedict et al. (2007) and Fouqu\'e et al. (2007). 
For the Magellanic Cepheids we have adopted periods, optical and NIR 
photometry from Laney \& Stobie (1994). The pulsation phases at which 
our stars were observed and selected characteristics are listed in 
Tables~2 and 3 for the Galactic and the Magellanic Cloud Cepheids, 
respectively.  
Distance and reddening estimates for Galactic Cepheids come from  
two different samples. The "Old Sample" includes 32 Cepheids (see columns 
4 and 5 in Table~2) and among them 25 objects have distance moduli provided 
by Storm et al. (2004, see their Table~3) and by Groenewegen et al. (2004, 
see their Table~3). 
These distances are based on the infrared surface brightness method and 
the two different calibrations provide, within the errors, the same 
distances. For the remaining seven objects distance estimates are not 
available in the literature. For five of them we determined the distance  
by combining the linear diameter from Laney \& Stobie (1995) with the 
$V,K$  unreddened magnitudes from Laney \& Stobie (1994), using two 
surface brightness-color calibrations:

\begin{itemize}
\item from Groenewegen (2004): we have combined Eq.~1 and Eq.~2  with 
Table~3 coefficients marked with the filled circle (the $V$ vs. $V-K$ relation)

\item from Fouqu\'e \& Gieren (1997): we have combined Eq.~1 with Eq.~27 
(the $V$ vs. $V-K$ relation)
\end{itemize}

The distance moduli derived with the two calibrations mentioned above agree 
very well (within 1\%) and we have adopted the distances determined with 
Groenewegen's calibration (these are the values listed in Table~2).
The "New Sample" includes 32 Cepheids (see columns 6 and 7 in Table~2) 
and among them 24 objects have distance moduli based on the infrared 
surface brightness method provided by Fouqu\'e et al. (2007, see their 
Table~7). The trigonometric parallaxes for l Car, $\beta$ Dor, 
and $\zeta$ Gem have been provided by Benedict et al. (2007). 
For EU Tau, we adopted the distance by Groenewegen et al. (2004), while 
for KQ Sco and SX Vel we adopted distances 
%Martin 
calculated by one of us (MG), following the general method outlined in 
Groenewegen (2007), but using the SB-relation and projection-factor 
from Fouqu\'e et al. (2007) for consistency.
The Cepheid AP Pup was not included in the analysis of the metallicity 
effect because an accurate estimate of its linear diameter is not available. 
For this object in Table~2 we only listed the apparent magnitudes. The same 
outcome applies to AX Vel, since Laney \& Stobie (1995) mentioned that 
the quality of the radius solution for this object was quite poor. 
Accurate reddening estimates for Galactic Cepheids have been recently
provided by Laney \& Stobie (2007), however, we typically adopted the 
reddening estimates used to determine individual Cepheid distances.

\begin{figure}
\begin{minipage}{0.45\textwidth}
\resizebox{\hsize}{!}{\includegraphics{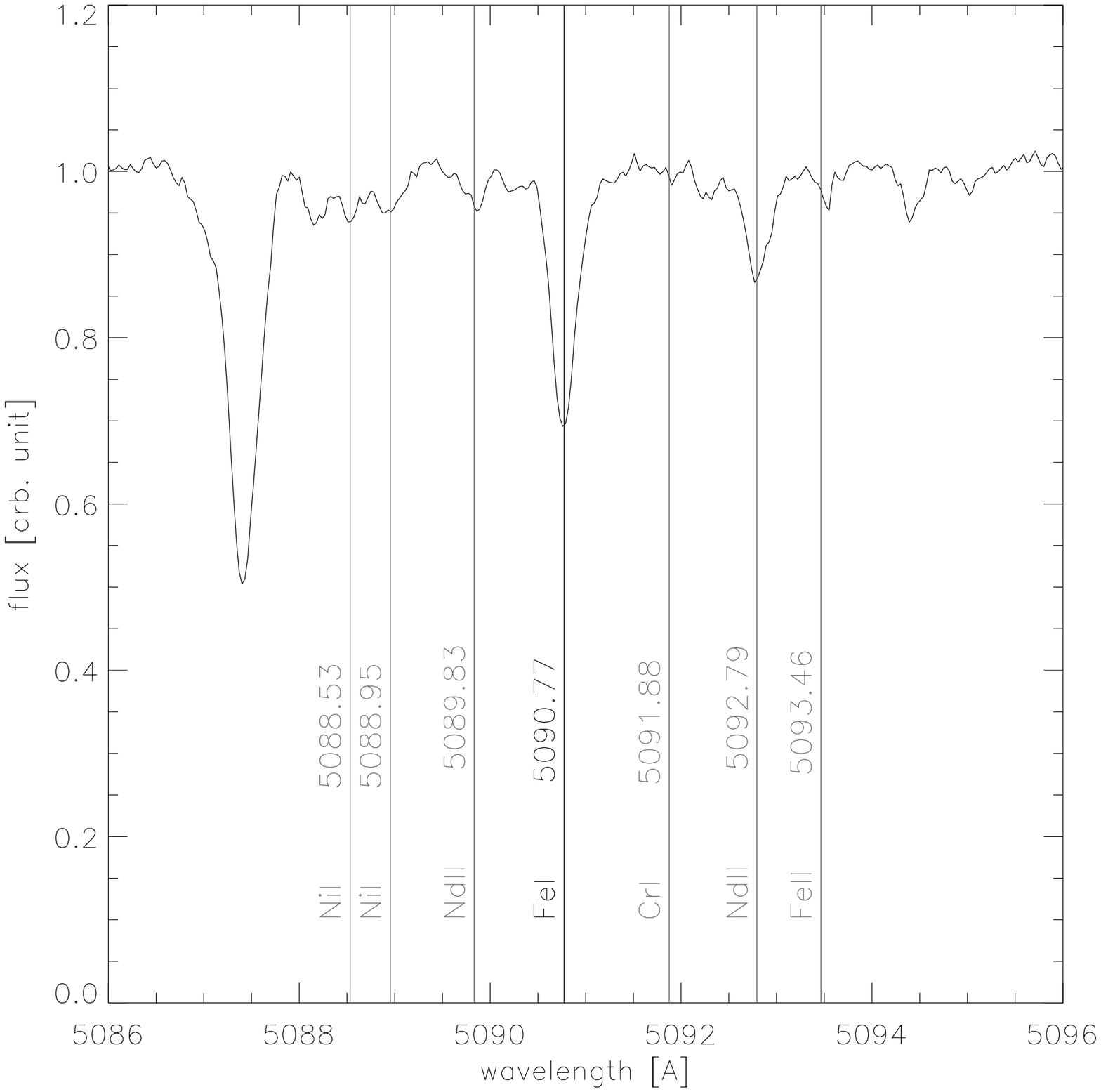}}
\end{minipage}
\begin{minipage}{0.45\textwidth}
\resizebox{\hsize}{!}{\includegraphics{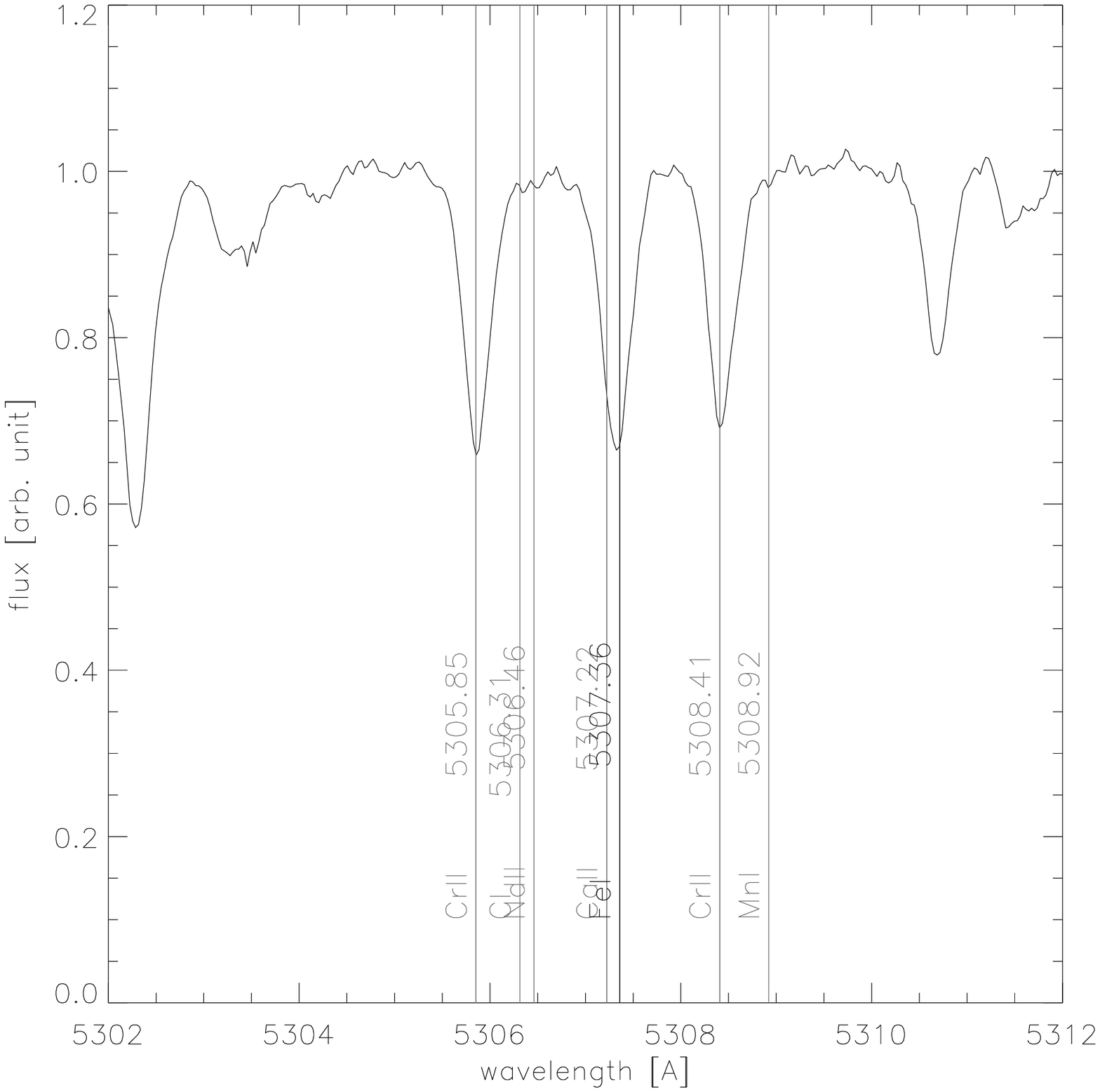}}
\end{minipage}
\caption{Examples of visual inspection on the selected iron lines. The iron lines are
plotted in black while the other elemental lines are plotted in gray. In the top 
panel it is shown an unblended line, while in the bottom panel there is an example 
of a blended line.}
\label{Fig2}
\end{figure}

The current Cepheid sample includes objects that are classified in 
the literature as fundamental pulsators, except EU Tau and SZ Tau, which are 
classified as first overtone pulsators, and AX Vel that is classified as 
one of the few double-mode pulsators in the Galaxy (Fernie et al. 1995; 
Sziladi et al. 2007). Indeed, when plotting our Cepheid sample in the 
log P vs $M_V$ plane, we confirm that EU Tau and SZ Tau are 
the only stars lying on the first overtone PL relation. Therefore, their 
observed periods have been "fundamentalised" using the relation
$P_0 = P_1/(0.716 - 0.027~log P_1)$ (Feast \& Catchpole 1997).

Approximately, 50\% of Galactic Cepheids in our sample are spectroscopic 
binaries (Szabados 2003) 
and one star ($\zeta$ Gem) is a visual binary (see last column of Table~2). 
According to current empirical evidence the companions are typically 
B and A-type main sequence stars, which are much less luminous (at least 3 
mag) than our main Cepheid targets. Only in the case of the two most luminous 
B dwarf, companions respectively of KN Cen and S Mus, we have detected a 
small contribution to the continuum level of the Cepheid spectra. Its effect 
on the final iron abundances will be discussed in Section 4.

\begin{figure}
   \centering
   \includegraphics[width=8.5cm]{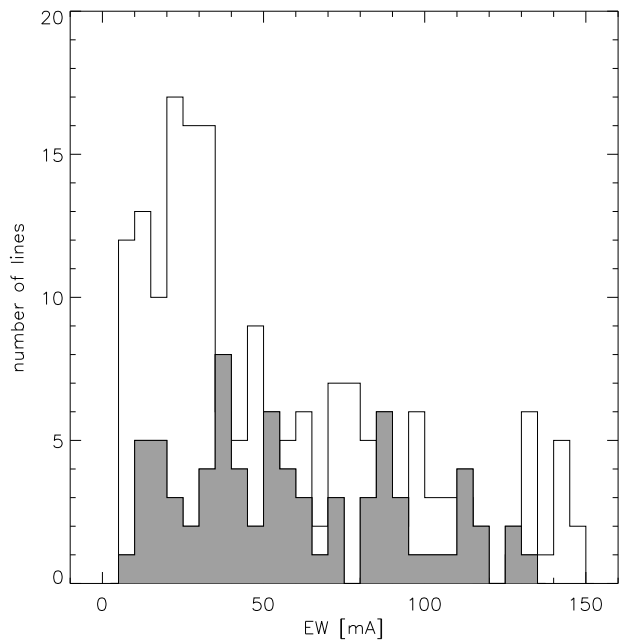}
      \caption{The distributions of the equivalent widths (EW) of our lines (empty histogram) 
and Fry \& Carney's lines (1997, dark grey histogram), in the case of the same star 
(VCen).}
         \label{Fig3}
   \end{figure}

\section{Methodology}

Below we present the methodology used to assemble the linelist and to determine 
the equivalent widths and the atmospheric parameters of our programme stars. 
All of these are key ingredients in the calculation of elemental abundances.

\subsection{Line list}
A crucial step of any spectral analysis in order to derive elemental abundances 
is a careful assembling of a line list. We have assembled our \ion{Fe}{i} - \ion{Fe}{ii} line list 
from Clementini et al.(1995), FC97, Kiss \& Vinko (2000) and Andrievsky et al.(2002a) 
plus a selection of lines from VALD (Vienna Atomic Line Database; Kupka et al. 1999; 
Ryabchicova et al. 1999).The VALD lines have been selected for effective temperatures 
typical of Cepheid stars (4500-6500 K). 
We have, then, visually inspected each line 
on the observed spectra, in order to check their profile and to discard blended lines. 
In order to do so, we have searched the VALD database (with the command extract 
stellar) for all the existing lines between 4800 and 7900 $\AA$ on stellar spectra 
characterized by stellar parameters typical of Cepheids (Teff=4500 K, 5500 K and 
6500 K, $\log g$=2 and vt=3km /s). We have, then, overplotted all the lines found in 
VALD, that fall within $\pm$ 3 $\AA$ from each of our iron lines, and checked for their 
possible contribution to the equivalent width of our iron line (an example is shown 
in Fig.~2). This test was carried out on 3 different observed spectra, characterized 
by effective temperatures (as found in the literature from previous analyses) close 
to 4500 K, 5500 K and 6500 K, respectively. Any contribution larger than 5\% of the 
line strength made us discard the iron line under scrutiny. There remains, of course, 
the possibility that some weak lines could be missing in the VALD compilation, 
but these are expected to be only weak lines, whose contribution would, then, be 
negligible.

Our final list includes 275 \ion{Fe}{i} lines and 37 \ion{Fe}{ii} lines, 
spanning the spectral range covered by our FEROS spectra. 
Our UVES spectra cover a narrower spectral range for which we can use 217 \ion{Fe}{i} 
lines and 30 \ion{Fe}{ii} lines. For all the lines, we have adopted the physical 
properties (oscillator strengths, excitation potentials) listed in VALD. Figure 
\ref{Fig3} shows the comparison between the distribution of the equivalent widths of 
our lines with those from FC97 for the Galactic Cepheid V Cen. As it can be 
clearly seen, our list is significantly larger and well samples the best range 
of equivalent widths, around 20 m$\AA$, as suggested by Cayrel (1988) to obtain 
reliable elemental abundances (iron in our case). The line list is presented in
Table~4, where the four columns list respectively the wavelength, the ion
identification, the excitation potential and the {\em $g_f$} values of 
each line. 

\begin{table}
\label{table:4}
\caption{Iron line list, from left to right the columns display wavelength, 
ion identification, excitation potential (EP) and {\em log gf} values. The 
table is available in its entirety via the link to the machine-readable 
version. A portion is shown here for guidance regarding its form and content.}
\centering
\begin{tabular}{c c c r} \hline\hline
$\lambda$ & $Ion$ & $EP$ & $\log gf$  \\ \hline
 4892.87  & \ion{Fe}{i}   &  4.22   &    -1.34 \\
 4917.23  & \ion{Fe}{i}   &  4.19   &    -1.29 \\
 4924.77  & \ion{Fe}{i}   &  2.28   &    -2.24 \\
 4932.08  & \ion{Fe}{i}   &  4.65   &    -1.49 \\
 4950.10  & \ion{Fe}{i}   &  3.42   &    -1.67 \\
 4973.11  & \ion{Fe}{i}   &  3.96   &    -0.95 \\
 4994.13  & \ion{Fe}{i}   &  0.91   &    -3.08 \\
 5029.62  & \ion{Fe}{i}   &  3.41   &    -2.05 \\
 5044.21  & \ion{Fe}{i}   &  2.85   &    -2.04 \\
 5049.82  & \ion{Fe}{i}   &  2.28   &    -1.36 \\
 5054.64  & \ion{Fe}{i}   &  3.64   &    -1.92 \\
 5056.84  & \ion{Fe}{i}   &  4.26   &    -1.96 \\
 5067.15  & \ion{Fe}{i}   &  4.22   &    -0.97 \\
 5072.67  & \ion{Fe}{i}   &  4.22   &    -0.84 \\
 5083.34  & \ion{Fe}{i}   &  0.96   &    -2.96 \\
\ldots    & \ldots        & \ldots  &  \ldots  \\ \hline
\hline
\end{tabular}
\end{table}

\subsection{Equivalent widths}

The second crucial step of our spectral analysis is the measurement of the 
equivalent widths (EW) of the iron lines we have assembled in our final list. 
At first, several independent and manual (i.e. with the IRAF splot task) 
measurements of the EW of the whole set of \ion{Fe}{i} and \ion{Fe}{ii} 
lines in a selected number 
of Cepheids were performed in order to test the reproducibility of our measures. 
However, because of the large number of lines and spectra, our final EW measurements 
were derived by using a semi-interactive routine (fitline) developed by one of us 
(PF, {\em fitline}). This code is based on genetic algorithms, which mimic how DNA 
can be affected to make the evolution of species (Charbonneau 1995). 
It uses a Gaussian fit, which is defined by four parameters: central 
wavelength, width, depth and continuum value of the line. A top-level view 
of the algorithm is as follows:

\begin{enumerate}

\item Compute an initial set of Gaussians, picking random values of the four parameter 
(scaled to vary between 0 and 1) and calculate the $\chi^{2}$ with the observed line for 
each Gaussian. 

\item Compute a new set of Gaussian from the 20 best fit of the previous ``generation'' 
introducing random modification in the values of the parameters (``mutation').

\item Evaluate the ``fitness'' of the new set ($\chi^{2}$ calculation for each Gaussian) 
and replace the old set with the new one.

\item Iterate the process (100 to 200 ``generations'') to get the best fit 
(lowest $\chi^{2}$) for each observed line.

\end{enumerate}

All the UVES spectra have been smoothed to improve the quality of the Gaussian fit 
(smooth\_step = 11 pix using {\em splot} task in IRAF). All the useful information 
from the spectra are preserved in the process. The selected iron lines, even on the 
smoothed spectra, do not show any contamination from other line. However, for about 
15\% of the lines, the Gaussian profile performed by {\em fitline}, could not satisfactorily 
reproduce the observed profile. The equivalent widths of these lines, usually 
very broad or asymmetric, were measured manually with the {\em splot} task. 
The mean difference, as computed for those lines for 
which both methods could be applied, is around 1.5 m$\AA$, comparable to the average 
error on the EW inferred from the quality of the data (Equation 7, Cayrel 1988). 
We can then safely use all our EW measurements, independently of the method we 
used to derive them.\\

For the determination of the metallicities, we have selected only lines with 
equivalent widths between 5 and 150 m$\AA$. The lower limit was chosen to be a 
fair compromise between the spectral characteristics and the need for weak 
lines for an optimal abundance determination. The upper limit was selected in 
order to avoid the saturated portion of the curve of growth. We note that a 
slightly higher upper limit (200 m$\AA$) was chosen for T Mon, SZ Tau and KQ Sco 
because these stars have very strong lines. This was done in order to keep the 
number of selected lines similar to the one used for the rest of the sample,
after checking that this higher upper limit does not have any effect on the 
final metallicities derived for these three stars.

Considering the mean difference on the EW obtained with {\em fitline} and {\em splot} 
and the uncertainty in the continuum placement from two measurements of the EW 
(carried out independently by two of us) we assume $\pm$ 3 m$\AA$ as error on the 
equivalent widths for the lines below 100 m$\AA$ and $\pm$ 5 m$\AA$ for the stronger 
features.

\subsection{Stellar parameters}

\begin{figure}
   \centering
   \includegraphics[width=8.5cm]{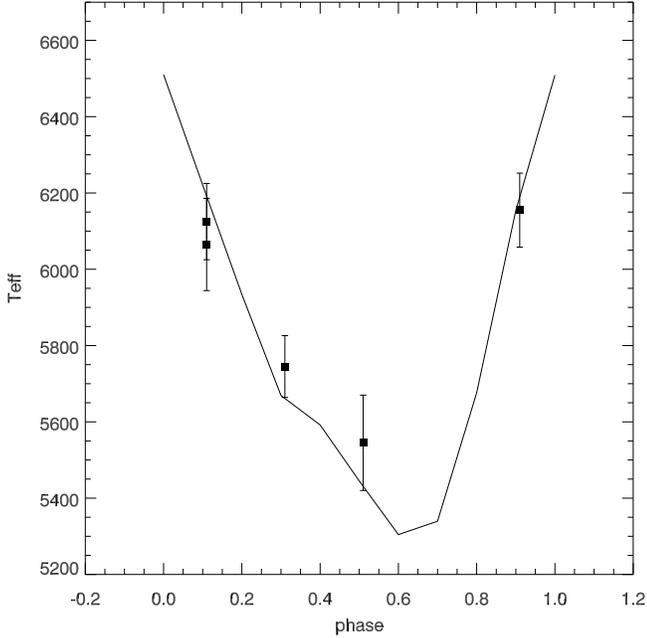}
      \caption{Comparison between the behaviour of the effective temperature of the Galactic 
Cepheid AP Puppis found by Pel (1978, solid line) and our results using the line 
depth ratios method for 5 phases (filled squares) with their associate errorbars.}
         \label{Fig4}
   \end{figure}

The stellar parameters were derived spectroscopically. We have determined the 
stellar effective temperature $T_{\rm eff}$ by using the line depth ratios method described 
in Kovtyukh \& Gorlova(2000). It is based on weak neutral metal lines (in our 
case, less than 100 m$\AA$ in equivalent width in order to exclude line broadening 
effects) with low excitation potentials, selected to obtain as close as possible 
pairs in wavelength space. Since the calibration of 
this method has been done using the FC97 scale of temperatures, our temperature 
scale is linked to theirs. The line depth ratios have the advantage of being 
independent of interstellar reddening and metallicity effects, uncertainties 
that instead plague other methods like the integrated flux method or colors 
calibrations (Gray 1994). The main uncertainties of the calibration of the line 
depth ratios, instead, lie in the accuracy of its zero point and slope, which 
have been thoroughly tested either with different colour-temperature relations 
or diameters measurements by Kovtyukh \& Gorlova (2000). As a sanity check 
on our temperature determinations, we have analysed five individual spectra 
of Galactic Cepheid AP 
Puppis taken at different phases along the pulsation period ($\phi$=0.11 - 
observed twice -, 0.31, 0.51, 0.91). Four out of these five spectra have
been collected in an independent observational campaign and the results 
have already been discussed by Lemasle et al. (2007). Figure \ref{Fig4} 
shows the effective temperatures derived for the five phases of AP Pup (filled 
squares) in comparison with the temperature variation estimated by Pel (1978) on 
the basis of accurate Walraven photometry. The agreement is indeed very good. 

The total number of line depth ratios adopted to estimate the temperature ranges from 
26 to 32 and from 20 to 28 for Galactic and Magellanic Cepheids, respectively. 
From this method, we have obtained effective temperatures with an intrinsic 
accuracy of about 40 K for the Galactic Cepheids and 50 K for the Magellanic 
ones (errors on the mean). Table 7 and Table 9 list our final effective temperatures 
for Galactic and Magellanic Cepheids, respectively.

Microturbulent velocity ({\em v$_{t}$}) and gravity ($\log g$) were constrained 
by minimizing the log([Fe/H]) vs EW slope (using the \ion{Fe}{i} abundance) and by 
imposing the ionisation balance, respectively. These two procedures are connected 
and require an iterative process (on average, between 5 and 7 iterations, depending 
on the star). We first achieved the minimization of the log([Fe/H]) vs EW slope and, 
subsequently, the ionisation balance. As first guess for the microturbulent 
velocity and the gravity, we adopted values typical of Cepheids ({\em v$_{t}$}=
3 km/s, $\log g$ = 2) as inferred from previous studies (FC97 and Andrievsky 
et al. 2002a, 2002b).

\begin{figure}
   \centering
   \includegraphics[width=8.5cm]{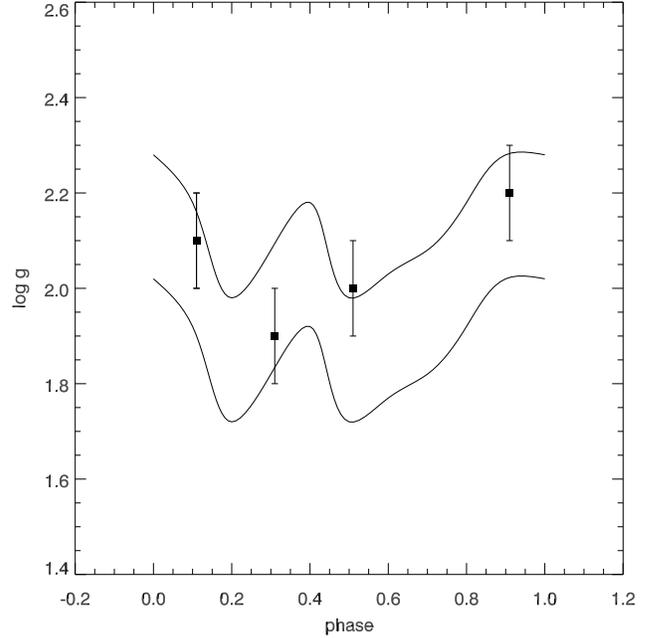}
      \caption{Comparison between the behaviour of the gravity of the Galactic Cepheid 
AP Puppis found by Pel (1978, the two solid lines indicate the upper and lower 
limits) and our results for 5 phases (filled squares) with their associate errorbars. 
At $\phi$=0.11, we have obtained the same value from the analysis of the two spectra.}
         \label{Fig5}
   \end{figure}

We first assumed the ionisation balance to be fulfilled when the difference 
between [\ion{Fe}{i}/H] and [\ion{Fe}{ii}/H] is less than the
standard error on [\ion{Fe}{ii}/H] (typically, 0.08-0.1 dex). However,we noticed 
that for most stars this condition was usually satisfied by more than one 
value of $\log g$, suggesting that our conditions for fulfillment might be 
too conservative and not very informative. We then checked which  $\log g$
value satisfies the ionisation balance within the standard error on [\ion{Fe}{i}/H] 
(typically ,0.02 dex). For 55 stars out of 68 we were able to reach
the \ion{Fe}{i}-\ion{Fe}{ii} balance within few hundredths of a dex. The corresponding 
$\log g$ values are the final gravity values quoted in Table 7 and Table 9. 
For the remaining 13 objects, we assumed as our final $\log g$ the value giving 
the "best" ionisation balance, i.e. the one with the smallest difference 
between [\ion{Fe}{i}/H] and [\ion{Fe}{ii}/H]. Also, it is worth noticing that for two 
stars (HV 2827and HV 11211) we had to increase the temperatures as 
determined from the line depth ratios by 50 K (which is still within the 
estimated error on the determination of $T_{\rm eff}$), in order to fulfill our 
requirements for a satisfactory ionisation balance.
Moreover, we note that the $\log g$ values derived for the five different 
phases of AP Puppis follow well the trend found by Pel (1978), as it can 
be seen in Fig. \ref{Fig5}. At $\phi$=0.11, we have obtained the same value 
from the analysis of the two spectra.

In order to determine the errors on the microtutbulent velocity and the 
gravity, we have run several iterations for each star, slightly modifying 
the values of these two quantities that fulfill the requirements mentioned 
above. We have estimated errors in $v_{t}$ to be 0.1 km/s and in $\log g$ to be 
0.10 dex.

\subsection{Model atmospheres}
We have derived the iron abundances of our stars by using the Kurucz WIDTH9 
code (Kurucz 1993) and LTE model atmospheres with the new opacity distribution 
functions (ODFs) computed by Castelli \& Kurucz (2003). These models neglect 
envelope overshooting. 
We have used a grid of solar metallicity models for the Galactic and LMC Cepheids 
and a grid of models computed assuming [Fe/H]=-1.0 and an $\alpha$-element enhancement 
of +0.4 dex for the SMC Cepheids (Gratton, Sneden \& Carretta, 2004). The grids of 
models have been interpolated in temperature in order to match the effective 
temperature derived for each star with the line depth ratio method and in 0.10 
dex gravity steps. 

Our choices of model atmospheres (in terms of ODFs, overshooting, 
metallicity, and $\alpha$-enhancement) have been thoroughly tested. A comparison between 
model atmospheres computed with the new ODFs (Castelli \& Kurucz 2003, hereafter 
identified as the 2003 models) and the old ODFs (Castelli et al 1997, hereafter 
identified as the 1997 models) shows small differences in the derived iron 
abundances: $\sim$ 0.01-0.03 dex. We tested the effect of different treatments of 
overshooting as implemented in different versions of Kurucz models by running the 
WIDTH9 code with two sets of the 1997 models computed with and without the 
``approximate'' overshooting. Differences in the final iron abundances amount to 
0.05 dex for Galactic and SMC Cepheids and around 0.06 dex for LMC Cepheids, 
with the models without overshooting giving the lower abundances.

Since the LMC has a mean metallicity around -0.3 dex,for this galaxy we also 
tested model atmospheres computed for [Fe/H]=-0.5 dex, finding differences of 
the order of $\sim$ 0.01-0.02 dex in the derived iron abundances compared to the 
solar metallicity models. Similar differences are found when the [Fe/H]=-0.5 
dex models are used also for the Small Magellanic Cloud (its mean metallicity 
is around -0.7 /-0 .8 dex). No differences in the derived iron content have been
found between models computed with and without the +0.4 dex $\alpha$-enhancement.

In all our computations, we have adopted a solar iron abundance of $log[n(Fe)]=7.51$ 
on a scale where $log[n(H)]=12$ (Grevesse \& Sauval, 1998) and we have assumed our 
final stellar metallicity to be the \ion{Fe}{i} value, which has been derived by a far 
larger number of lines with respect to \ion{Fe}{ii}.

\begin{figure}
   \centering
   \includegraphics[width=8.5cm]{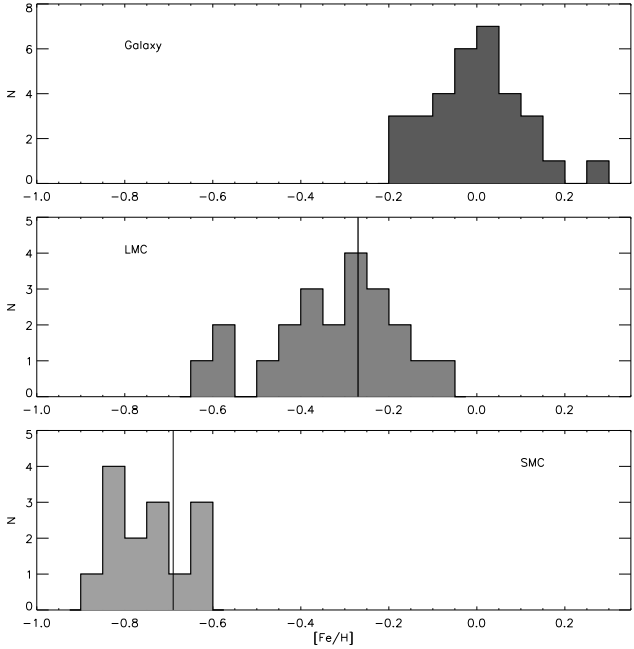}
      \caption{Histograms of the [Fe/H] ratios derived for all the stars of 
our sample in the Galaxy, the LMC and the SMC. The solid lines indicate 
the mean values found using F and K supergiants by Hill et al. (1995) 
and Hill (1997) (see Table 8 and Section 4.3). We found a good agreement 
with the results obtained for non-pulsating stars of the same age of Cepheids.}
         \label{Fig6}
   \end{figure}

\section{Iron abundances}

Our final iron abundances, together with the adopted stellar parameters, 
are listed in Table 7 and Table 9 for the Galactic and Magellanic Cepheids, 
respectively. 

For our Galactic sample, we find that the mean value of the iron 
content is solar ($\phi$ = 0.10, see Fig.~6), with a range of values between -0.18 
dex and +0.25 dex. These two extremes are represented respectively by 2 and 1 
stars (out of the 32 in total). Our Galactic Cepheids span a narrow range of 
Galactocentric distances (the mean Galactocentric distance of our sample is 
7.83 $\pm$ 0.88 kpc, see Fig.~7), thus preventing us from giving any indications 
about the metallicity gradient in the Galactic disc.

For the LMC sample, we find a mean metallicity value of $\sim$ -0.33 dex ($\phi$ = 
0.13, see Fig.~6), with a range of values between -0.62 dex and -0.10 dex. Here, 
the more metal-rich 
extreme is just an isolated case, while the metal-poor end of the distribution is 
represented by 3 stars.

For the SMC sample, we find that the mean value is about $\sim$ -0.75 dex ($\phi$ 
= 0.08, see Fig.~6), with a range of values between -0.87 and -0.63.

As we have already mentioned in Section 2, there are two binary stars 
(KN Cen and S Mus) among our Galactic sample with a bright B dwarf 
as a companion. In the spectral range covered by our spectra, these 
bright companions give a contribution only to the continuum level, 
because all their absorption lines fall in the ultra-violet region 
(these are very hot stars, with effective temperature $\sim$ 20,000 K). 
In order to test the effect of this contribution, we have subtracted 
from the observed spectrum of the two binaries the estimated optical 
spectrum of the B dwarf, which only consists of a continuum without any 
line. On the resulting spectra, which we assume to be the true spectra 
of our Cepheids, we have then remeasured the EWs of a sub-set of iron lines 
and derived the metallicity. In the case of KN Cen the differences in the 
EWs measured on the observed spectrum and the true Cepheid spectrum are 
negligible (i.e. within the errors), thus the iron content we have derived 
for this star is robust. Regarding S Mus, instead, the EWs measured on the 
true spectrum are 12-15\% larger than the ones measured on the spectrum we 
have observed. Thus, the true iron abundance of this Cepheid star should 
be 0.1 dex higher than the one we have derived. This happens because the 
intensity of the lines (due only to the Cepheid contribution) when compared 
to the continuum of the combined spectra is, in percentage, less than 
the intensity of the lines compared to the continuum of the Cepheid alone. 
In other words the contribution of the companion makes the lines of the 
Cepheid weaker. Please note that the iron content reported for S Mus in 
Table 7 has already been corrected for the above mentioned effect.

\begin{figure}[b]
   \centering
   \includegraphics[width=8.5cm]{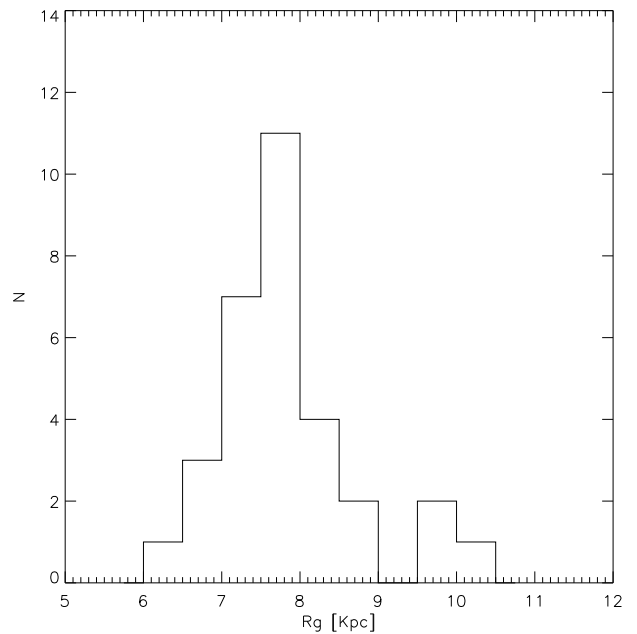}
      \caption{Histograms of the Galactocentric distances derived for all the Galactic 
Cepheids with individual heliocentric distances.}
         \label{Fig7}
   \end{figure}

\begin{table}
\label{table:5}
\caption{Effects on measured $\ion{Fe}{i}$ and $\ion{Fe}{ii}$ abundances 
caused by changes in atmospheric parameters.}
\begin{center}
\begin{tabular}[b]{ccc}\hline \hline
   & $\Delta$[\ion{Fe}{i}/H]    & $\Delta$[\ion{Fe}{ii}/H]\\ \hline
$\Delta T_{\rm eff}$=+100 K     & +0.07 dex & +0.00 dex \\
$\Delta \log g=$+0.1 dex         & +0.00 dex & +0.04 dex \\
$\Delta$ {\em v$_t$}=+0.1 Km/s  & -0.03 dex & -0.02 dex \\ \hline
\hline
\end{tabular}
\end{center}
\end{table}

\subsection{Uncertainties}

The internal uncertainties in the resulting abundances are due to errors 
in the atomic data ($gf$-values) and EW measurements.
We have estimated, on average, an internal error in the [Fe/H] 
determination of 0.10 dex.

It is also important to understand the effects of potential systematic 
errors in the stellar parameters on the final derived abundances 
(see Table 5). In order to do so we have determined curves of growth 
for different effective temperatures, microturbulent velocities and 
gravities for both \ion{Fe}{i} and \ion{Fe}{ii}. As expected, we find 
that  \ion{Fe}{i} abundances marginally depend on the gravity, 
whereas \ion{Fe}{ii} does not depend on the effective temperature. 
An increase in temperature of 100~K, at fixed {\em v$_{t}$} and $\log g$, 
results in an increase in [\ion{Fe}{i}/H] 
of about 0.07 dex. An increase in {\em v$_{t}$} of 0.1 km/s gives a decrease 
in [\ion{Fe}{i}/H] of about 0.03 dex, for constant $T_{\rm eff}$ and 
$\log g$, and we obtain a decrease of 0.02 dex for [\ion{Fe}{ii}/H]. An increase 
in $\log g$ of about 0.1 dex produces an increase in [\ion{Fe}{ii}/H] of about 
0.04 dex (with fixed {\em v$_{t}$} and $T_{\rm eff}$). 

An additional potential source of uncertainty 
and concern comes from the fact that our metallicities have been 
derived from single epoch observations. However, we note that FC97 
did not find any significant difference in their derived [Fe/H] as 
a function of phase (the test was performed on four of their longest 
period cepheids) and Luck \& Andrievsky (2004) and Kovtyukh et al. (2005) 
show that the elemental abundances for Cepheids with a period between 6 
and 68 days are consistent for all pulsation phases. Moreover, our 
exercise on AP Puppis (see section 3.3 and Table 6) further confirms 
these conclusions, for a star even with a shorter period (5 days). 
As it can be seen in Table 6, not only do the derived stellar parameters nicely 
follow the expectations from photometry, as shown in Fig. \ref{Fig4} and 
Fig. \ref{Fig5}, but the metallicities measured at all phases agree within 
the errors ($\sim 0.1$ dex).

\begin{table}
\label{table:6}
\caption{Stellar parameters and iron abundances along the pulsation cycle  
of the Galactic Cepheid AP Puppis. Phase 0.11 was observed twice 50 days 
apart, leading to very consistent results.}
\begin{minipage}[b]{\columnwidth}
\begin{center}
\begin{tabular}[b]{ccccc}\hline \hline
Phase & $T_\mathrm{eff}$ & {\em v$_{t}$} & $\log(g)$ & $[Fe/H]$ \\\hline
0.11  & 6070 & 3.05 & 2.1 & -0.07 \\
0.11  & 6130 & 3.00 & 2.1 & -0.03 \\
0.31  & 5750 & 2.80 & 1.9 & -0.03 \\
0.51  & 5550 & 3.30 & 2.0 & -0.06 \\
0.91  & 6160 & 4.20 & 2.2 & -0.13 \\\hline
\hline
\end{tabular}
\end{center}
\end{minipage}
\end{table}

\begin{table*}
%\begin{minipage}[b]{\columnwidth}
\label{table:7}
\caption{Stellar parameters and \ion{Fe}{i} and \ion{Fe}{ii} abundances of our Galactic sample. 
The \ion{Fe}{i} values have been adopted as final [Fe/H]. We compare our results, where 
it is possible, with previous investigations: Fry \& Carney (1997); Andrievsky et al. 
(2002a,b,c), Luck et al. (2003).}
\begin{center}
\renewcommand{\footnoterule}{}
\begin{tabular}{l c c c c c c c} \hline\hline
ID & $T_{\rm eff}$ & $v_{t}$ & $\log g$ & [\ion{Fe}{i}/H] & [\ion{Fe}{ii}/H]& [Fe/H]$_{FC}$ & [Fe/H]$_{A}$   \\ \hline

  l  Car    & 4750 & 3.60 & 0.4 &  0.00 &  0.00 &  \ldots    &   \ldots	     \\
  U  Car    & 5980 & 3.00 & 1.2 & +0.17 & +0.13 &  \ldots    &   \ldots	     \\
  V  Car    & 5560 & 3.05 & 1.8 & -0.07 & -0.02 &  \ldots    &   \ldots      \\
 WZ  Car    & 4520 & 1.80 & 2.1 & +0.08 & +0.19 &  \ldots    &   \ldots	     \\
  V  Cen    & 6130 & 2.80 & 1.9 & +0.04 & -0.03 &   -0.12    &  +0.04$^{a}$  \\
 KN  Cen    & 5990 & 3.80 & 1.6 & +0.17 & +0.09 &  \ldots    &  \ldots	     \\
 VW  Cen    & 5240 & 4.20 & 1.2 & -0.02 & +0.07 &  \ldots    &  \ldots	     \\
 XX  Cen    & 5260 & 2.95 & 1.3 & +0.04 & +0.04 &  \ldots    &  \ldots	     \\
$\beta$ Dor & 5180 & 3.00 & 1.3 & -0.14 & -0.11 &  \ldots    &  -0.01$^{a}$  \\
$\zeta$ Gem & 5180 & 3.70 & 1.4 & -0.19 & -0.14 &   -0.04    &  +0.06$^{a}$  \\
 GH  Lup    & 5480 & 3.60 & 1.5 & +0.05 & +0.01 &  \ldots    &  \ldots	     \\
  T  Mon    & 4760 & 3.80 & 0.6 & -0.04 & -0.03 &   +0.09    &  +0.21$^{b}$  \\
  S  Mus    & 5780 & 2.75 & 1.8 & +0.13 & +0.19 &   \ldots   &  \ldots	     \\
 UU  Mus    & 5900 & 3.05 & 1.7 & +0.11 & +0.05 &   \ldots   &  \ldots	     \\
  S  Nor    & 5280 & 2.80 & 1.1 & +0.02 & +0.04 &   -0.05    &  +0.06$^{a}$  \\
  U  Nor    & 5230 & 2.60 & 1.1 & +0.07 & +0.13 & \ldots     &  \ldots	     \\
  X  Pup    & 5850 & 3.30 & 1.4 & +0.04 & -0.05 & \ldots     &   0.00$^{a}$  \\ 
 AP  Pup    & 6070 & 3.05 & 2.1 & -0.07 & -0.07 & \ldots     & \ldots        \\
 AQ  Pup    & 5170 & 3.10 & 0.8 & -0.07 & -0.09 & \ldots     &  -0.14$^{b}$  \\
 BN  Pup    & 5050 & 2.95 & 0.6 & -0.10 & -0.07 & \ldots     &  +0.01$^{c}$  \\
 LS  Pup    & 6550 & 3.50 & 2.2 & -0.16 & -0.10 & \ldots     &  \ldots	     \\
 RS  Pup    & 4960 & 3.50 & 0.7 & +0.09 & +0.10 & \ldots     &  +0.16$^{b}$  \\
 VZ  Pup    & 5230 & 3.25 & 1.1 & -0.17 & -0.15 & \ldots     &  -0.16$^{c}$  \\
 KQ  Sco    & 4840 & 3.60 & 0.7 & +0.21 & +0.27 & \ldots     &  +0.16$^{d}$  \\
 EU  Tau    & 6060 & 2.30 & 2.1 & +0.04 & +0.02 & \ldots     &  -0.03$^{a}$  \\
 SZ  Tau    & 5880 & 2.80 & 1.7 & +0.07 & +0.04 &  -0.02     &  +0.06$^{a}$  \\
  T  Vel    & 5830 & 2.55 & 1.8 & +0.10 & +0.03 & \ldots     &  -0.04$^{b}$  \\
 AX  Vel    & 5860 & 3.10 & 1.8 & +0.02 & -0.06 & \ldots     &  \ldots	     \\
 RY  Vel    & 5250 & 4.10 & 1.2 & -0.07 & -0.06 & \ldots     &  -0.03$^{b}$  \\
 RZ  Vel    & 5140 & 4.40 & 1.6 & -0.19 & -0.14 & \ldots     &  -0.11$^{b}$  \\
 SW  Vel    & 6590 & 3.75 & 1.7 & -0.24 & -0.17 & -0.10      &  -0.05$^{b}$  \\
 SX  Vel    & 5380 & 2.55 & 1.2 & +0.02 & -0.02 & \ldots     &  -0.04$^{b}$  \\ \hline
\hline
\multicolumn{8}{l}{$^{a}$ Luck et al. (2003).}\\    
\multicolumn{8}{l}{$^{b}$ Andrievsky et al. (2002a).}\\
\multicolumn{8}{l}{$^{c}$ Andrievsky et al. (2002c).}\\
\multicolumn{8}{l}{$^{d}$ Andrievsky et al. (2002b).}\\
\end{tabular}
\end{center}
\end{table*}

%%%%%%%%%%%%%%%%%%%%%%%%%%%%%%%%%%%%%%%%%%%%%%%%%%%%%%%%%%%%%%%%%%
\subsection{Comparison with previous studies on Cepheids}

Before we address the key issue concerning the metallicity dependence 
of the Cepheid PL relation (cf. Sect. 5), let us first compare our results 
to previous works. For this purpose, we have selected the chemical analyses 
of FC97, Andrievsky et al. (2002a, 2002b, 2002c) and Luck et al. (2003) 
for the Galactic Cepheids and LL92 and L98 for the Magellanic Clouds. 
When necessary, we have rescaled the literature [Fe/H] values to the solar 
iron abundance we have adopted ($log[n(Fe)]=7.51$). One should keep in mind 
that the same stars have likely been observed at different phases and 
analysed with different tools and model atmospheres, which may lead to 
the determination of different combinations of stellar parameters. 
In the case of observations at 
different phases, as we already mentioned in the previous section, our results 
on the five spectra of AP Pup and the conclusions of Luck \& Andrievsky (2004) 
and Kovtyukh et al. (2005) show that the elemental abundances do not depend on 
the phase. When multi-phase observations of the same star were available in the 
literature (this is the case for some of the Galactic Cepheids), we have compared 
the [Fe/H] values obtained for the same phase of our observations. Otherwise, 
we chose the iron content corresponding to the closest value of effective 
temperature to ours. Regarding the determination of different sets of stellar 
parameters, we do not have the capabilities (in terms of analytical codes and 
different sets of model atmospheres) in order to properly take this into account. 
However, we did signal such differences whenever we noted them.

In total, for the Galactic sample, we have 6 stars in common with FC97 and 18 
with the entire sample of Andrievsky's group, while for the MC sample, we have 
3 stars in common with LL92 and 7 with L98.

\subsubsection{Galactic Cepheids}

The mean difference between our results and those of FC97 and Andrievsky analyses is 
comparable to the difference between FC97 and Andrievsky's values (0.08 $\pm$ 0.02). 
In more detail, the mean difference between our iron abundances and FC97 is 
0.09 $\pm$ 0.02, which is satisfactory. For 4 stars (V Cen, S Nor, TMon, $\zeta$ Gem) 
the agreement is at 1$\sigma$ level, for the remaining 2 stars (we have 6 in common, 
in total) the agreement is well within the quoted uncertainties. We note that 
our derived abundances for V Cen and S Nor are more in agreement with the 
metallicity derived by Andrievsky et al. than with FC97. 

When comparing our 
results to Andrievsky's, we obtain a mean difference of 0.07 $\pm$ 0.05. For 14 
stars (out of 18), the iron abundances agree quite well within the associated 
errors. Of the remaining 4 stars, we note that SW Vel, $\beta$ Dor, and T Mon agree 
within the standard deviation ($\sigma$=0.20, 0.19, 0.22 dex respectively) instead 
of the standard error, that is the condition suggested by Kovtyukh et al. 
(2005a) for the independence of the elemental abundance on the phase. This is 
not the case for $\zeta$ Gem, for which we cannot explain the difference (0.20 dex) 
but for which we find instead an agreement with FC97.

\subsubsection{Magellanic Cepheids}

In order to properly compare our results for the Magellanic Cepheids with the 
values obtained by LL92 and L98, we first had to rescale the latter values for 
the difference in the adopted solar iron abundance between us (7.51) and them 
(7.67 and 7.61, respectively). These revised values are listed in the last column 
of Table 9.

In general, the mean metallicities that L98 found with their complete sample (-0.30 
dex and -0.74 dex for the LMC and SMC, respectively) are in very good agreement 
with our results (-0.33 dex and -0.75 dex). They also found a similar spread in 
iron for the MCs. With LL92, instead, there is a good agreement in the case of LMC but 
a difference for the metallicity of the SMC (0.15 dex greater than our mean value). 
Our derived abundances are always smaller than the values derived by LL92. However, 
the number of objects in common is too small to constrain the effect on a quantitative 
basis.

When we move to a star-by-star comparison, larger differences emerge. 
The comparison with L98 abundances discloses a very good agreement for one object 
(HV 837) and a plausible agreement, i.e. within the standard deviation error, for 
other 3 stars (HV 5497, HV 824 and HV 834). Regrettably, this is not the case of 
HV 879, HV 2827 and HV 829 for which we note rather large discrepancies ($\sim$ 0.2 
- -0.3 dex) that remain unexplained. However, we note that L98 used different 
analytical codes, different oscillator strengths, and different values of 
the stellar gravity (they adopted the "physical" gravity calculated from the 
stellar mass, the luminosity and the temperature). A combination of all these 
three factors could well account for the observed differences.

\begin{table}
\label{table:8}
\caption{Comparison of the mean metallicities of the Magellanic Clouds with 
previous studies. The number of studied stars are also listed. RB89: Russell \& 
Bessell (1989), RD92: Russell \& Dopita(1992), R93: Rolleston et al. (1993), 
H95: Hill et al. (1995), H97: Hill (1997), K00: Korn et al. (2000), A01: 
Andrievsky et al. (2001), R02: Rolleston et al. (2002),  
G06: Grocholski et al. (2006), T07: Trundle et al. (2007), 
P08: Pompeia et al. (2008).}
\begin{minipage}[b]{\columnwidth}
\begin{center}
\begin{tabular}[b]{cccl}\hline \hline
Reference  & [Fe/H]$_{LMC}$ & [Fe/H]$_{SMC}$     & Notes \\\hline
This work  & -0.33$\pm$0.13 & -0.75 $\pm$ 0.08   & 22+14 Cepheids    \\
RB89       & -0.30$\pm$0.20 & -0.65 $\pm$ 0.20   & 8+8 F supergiants \\
R93        & 	 \ldots	    & -0.80 $\pm$ 0.20\footnote[1]{Mean metallicity based on oxygen.} &   4 B stars	     \\
H95        & -0.27$\pm$0.15 &     \ldots         &   9 F supergiants \\
H97        & 	  \ldots	    & -0.69 $\pm$ 0.10   & 6+3 K supergiants \\
K00        & -0.40$\pm$0.20 & -0.70 $\pm$ 0.20   &   6 B stars	    	\\
A01        & -0.40$\pm$0.15 &      \ldots	         &   9 F supergiants	\\
R02        & -0.31$\pm$0.04 &      \ldots	         &   5 B stars      	\\
G06        & -0.3 -- -2.0\footnote[2]{Metallicity based on calcium triplet.}&      \ldots                & 200 giants\\  
T07        & -0.29$\pm$0.08 &      -0.62$\pm$0.14 &  13+5 cluster giants  	\\
P08        & -0.3 -- -1.8&      \ldots	         &   59 red giants     	\\\hline
\end{tabular}
\end{center}
\end{minipage}
\end{table}
As already mentioned, larger differences (0.4 - 0.6 dex) are found with 
respect to LL92, for the 3 stars we have in common.  
Following the referee suggestion we performed a more quantitative comparison
for the three Cepheids in common with LL92, namely HV~865, HV~2195 and HV~2294.
The calibrated spectra were kindly provided in electronic form by R.E. Luck.
By adopting the same atmospheric parameters used by LL92 we find the following
iron abundances: [\ion{Fe}{i}/H]= -0.75 (HV~865), -0.60 (HV~2195) and -0.15
(HV~2294). The comparison with the iron abundances provided by LL92 (see the
last column in Table~9) indicates that the new measurements are $\sim 0.3$ dex
systematically more metal-poor. They also agree, within the errors, reasonably 
well with current measurements (see column 5 in Table~9). Thus supporting
the arguments quoted above to account for the difference between the two
different iron measurements. 
Moreover, one should keep in mind that the quality of the LL92 data set 
is significantly lower than ours (R $\approx$ 18,000
vs 40,000) and that their oscillator strengths may differ from our selection: 
they were taken from the critical compilations of Martin, Fuhr \& Wiese (1988) 
and Fuhr, Martin \& Wiese (1988), which are included in VALD but not among the 
references used for our gf-values. Moreover, they have used different analytical 
tools: MARCS model atmospheres (Gustafsson et al. 1975) and a modified version 
of the LINES line-analysis code and MOOG (Sneden 1973) synthesis code. They 
also state that their method overestimates the equivalent widths of the weak 
lines.

\subsection{A comparison with different stellar populations in the Magellanic 
Clouds}

The mean metallicities of our Magellanic sample are in good agreement with 
previous results obtained for F, K supergiants and B stars in the Magellanic 
Clouds (see Table 8).

Russell \& Bessell (1989) found a mean [Fe/H]= -0.30 $\pm$ 0.2dex in the 
LMC and [Fe/H]= -0.65 $\pm$ 0.2 dex in the SMC, analysing high-resolution 
spectra of 
16 F supergiants (8 for each galaxy). In 1995 Hill et al. obtained, from 
9 F supergiants from the field of the LMC, a mean iron abundance of -0.27 $\pm$
0 .15 dex and Hill (1997) found a mean [Fe/H]= -0.69 $\pm$ 0.1 dex, analysing 
6 K supergiants in the SMC. Andrievsky et al.(2001) re-analysed the 
sample of F LMC supergiants studied by Hill et al. (1995) and obtained 
a slightly lower mean value: [Fe/H]= -0.40 $\pm$ 0.15 dex.

Regarding the B type stars, Rolleston et al. (1993) and Rolleston et al. 
(2002) found, from the analysis of 4 stars in the SMC and 5 stars in 
the LMC, respectively, mean values of metallicity of -0.8 $\pm$ 0.20 dex 
and -0.31 $\pm$ 0.04 dex. Korn et al.(2000) obtained for the LMC 
(from 6 B stars) a mean [Fe/H]= -0.40 $\pm$ 0.2 dex and [Fe/H]=-0.70$\pm$0.2 dex 
for the SMC (from 3 B stars). Grocholski et al. (2006) using homogeneous calcium 
triplet measurements for 200 stars belonging to 28 different LMC clusters covering 
a broad range of cluster ages, found by transforming the Ca abundance in iron 
abundance that the metallicity ranges from [Fe/H]$\sim$-0.3 to [Fe/H]$\sim$-2.0.  
More recently, Trundle et al. (2007) collected high resolution spectra with 
FLAMES@VLT for a good sample of Magellanic giants and found a mean metallicity 
of [Fe/H=$-0.62 \pm 0.14$, 5 giants in the SMC cluster NGC~330, and 
of [Fe/H]=$-0.29 \pm 0.08$, 13 giants in the LMC cluster NGC~2004.  
By using similar data for 59 red giant stars in the LMC inner disk, 
Pompeia et al. (2008) found that their metallicity ranges from [Fe/H]$\sim$0.0 
to [Fe/H]$\sim$-1.80. 

In conclusion, the mean iron content of Cepheids, for the Magellanic Clouds, 
agree very well with the results obtained for non variable stars of similar 
age and stars that are Cepheid's progenitors. Cepheids do not show any difference 
with these two other populations.

\begin{figure*}
   \centering
   \includegraphics[width=12cm]{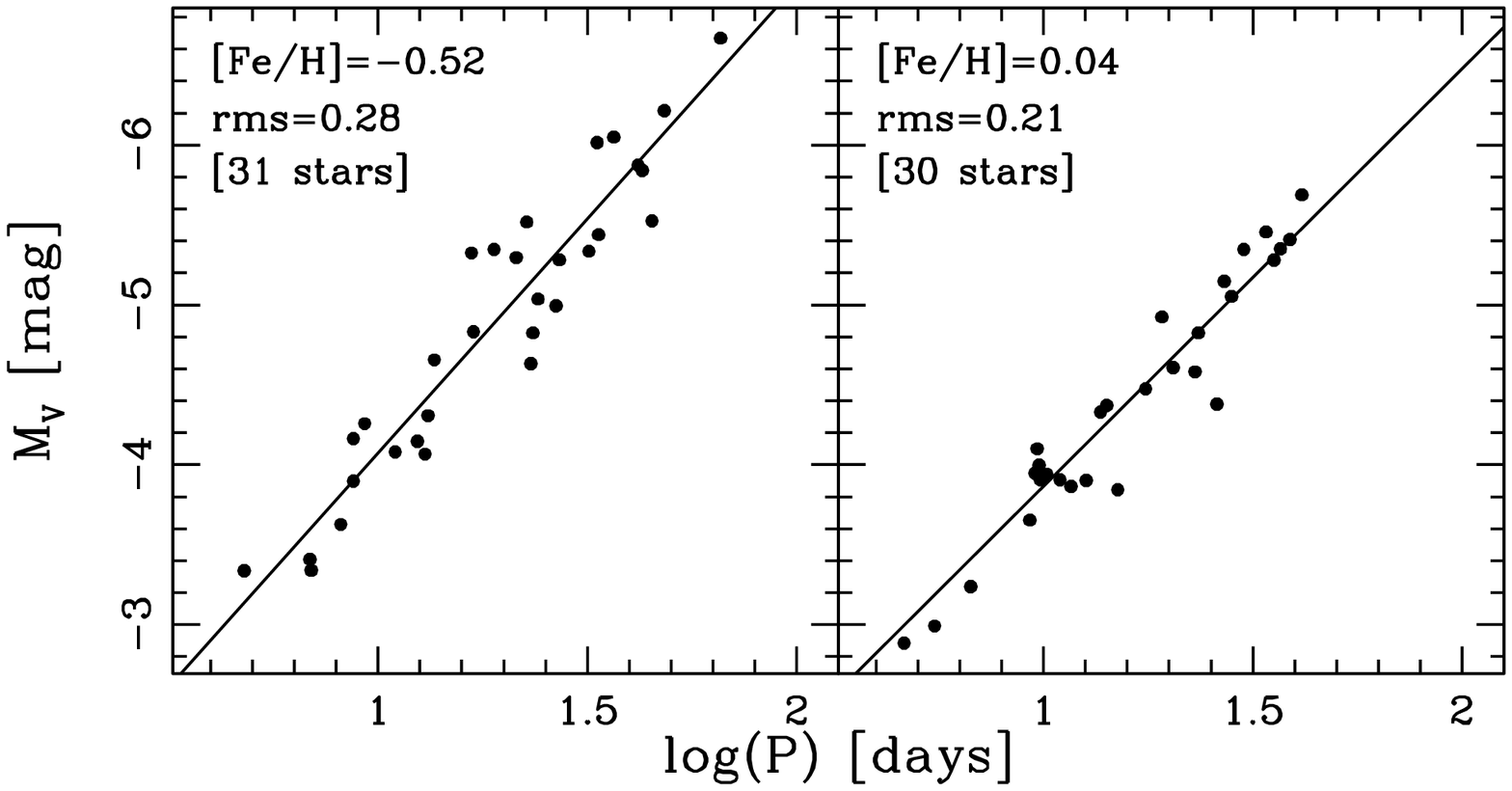}
   \centering
   \begin{tabular}{cc}
      \resizebox{80mm}{!}{\includegraphics{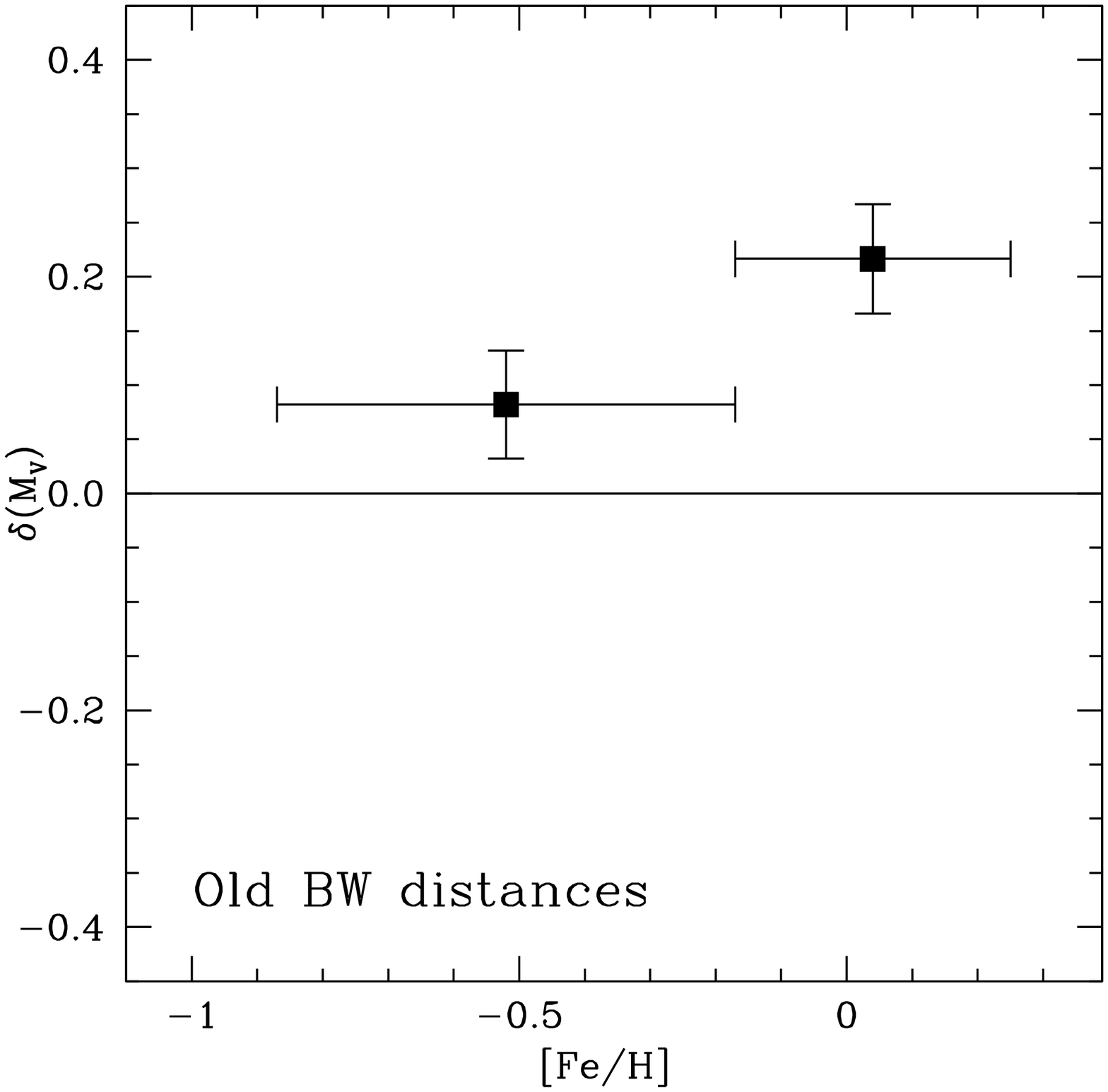}} &
      \resizebox{80mm}{!}{\includegraphics{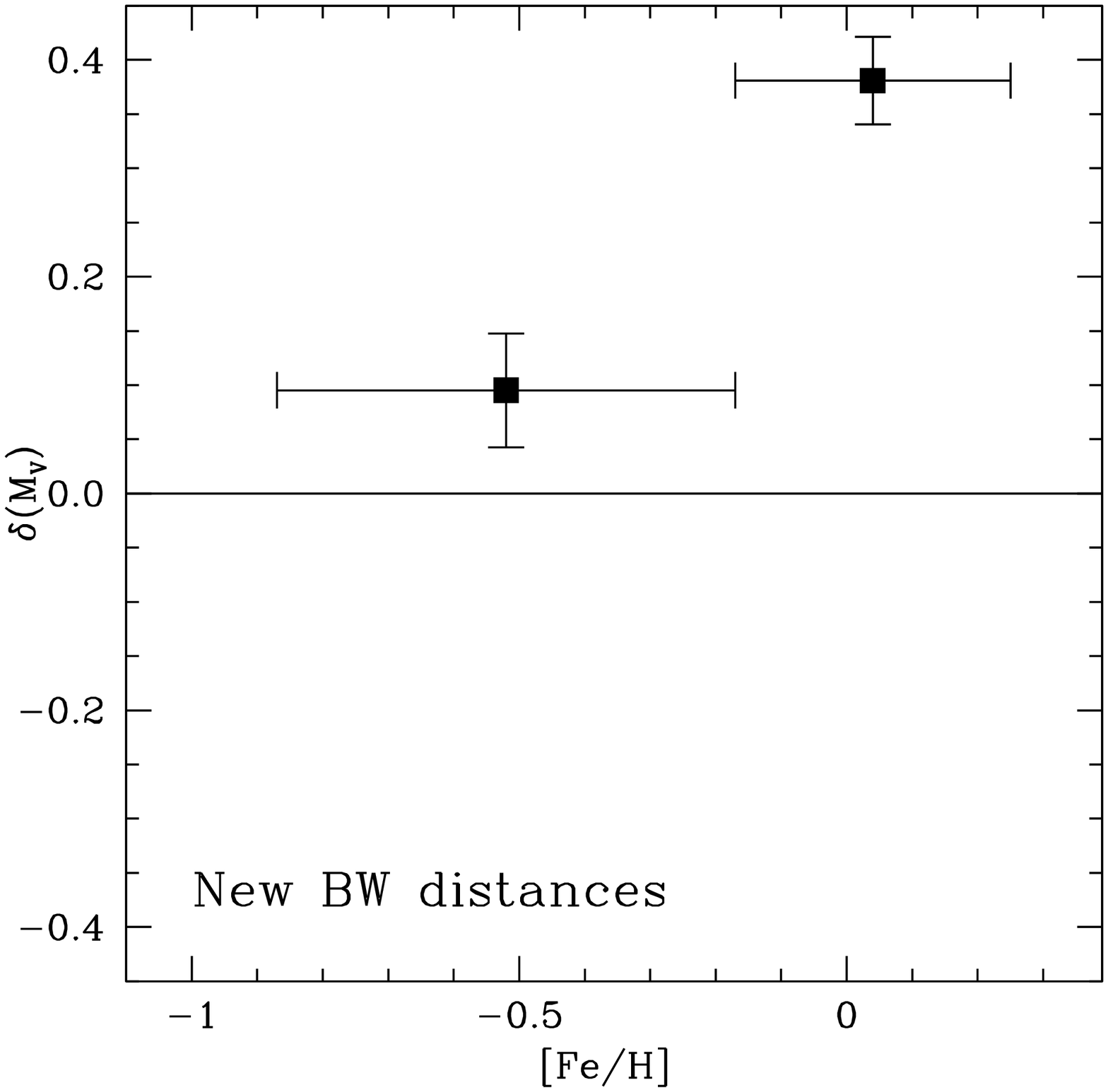}} \\
   \end{tabular}
      \caption{The top panels show the PL relations calculated in each bin for the $V$ band. 
The bottom panels show the residuals $\delta(M_V)$ as a function of the iron abundance
for both the "Old" and the "New" Galactic Cepheid distances. The mean values of $\delta(M_V)$ 
in each metallicity bin are plotted as filled squares, with the vertical error-bars 
representing its associated error. The horizontal bars indicate the dimension of the bins. 
The horizontal solid lines indicate the null value which corresponds to an independence of 
the PL relation from the iron content.}
         \label{Fig8}
   \end{figure*}

\begin{figure*}
   \centering
   \includegraphics[width=12cm]{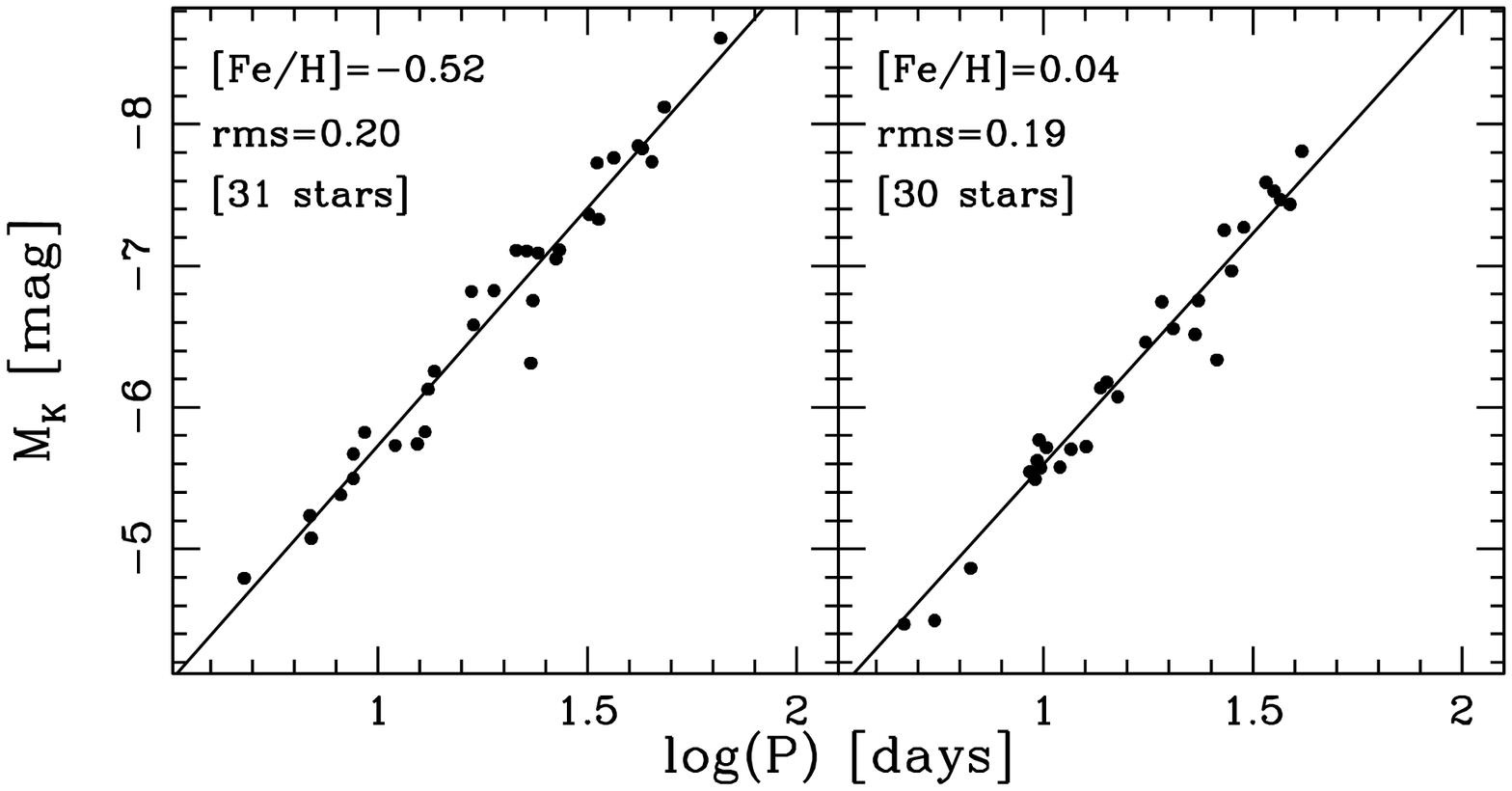}
   \label{Fig:9}
   \centering
   \begin{tabular}{cc}
      \resizebox{80mm}{!}{\includegraphics{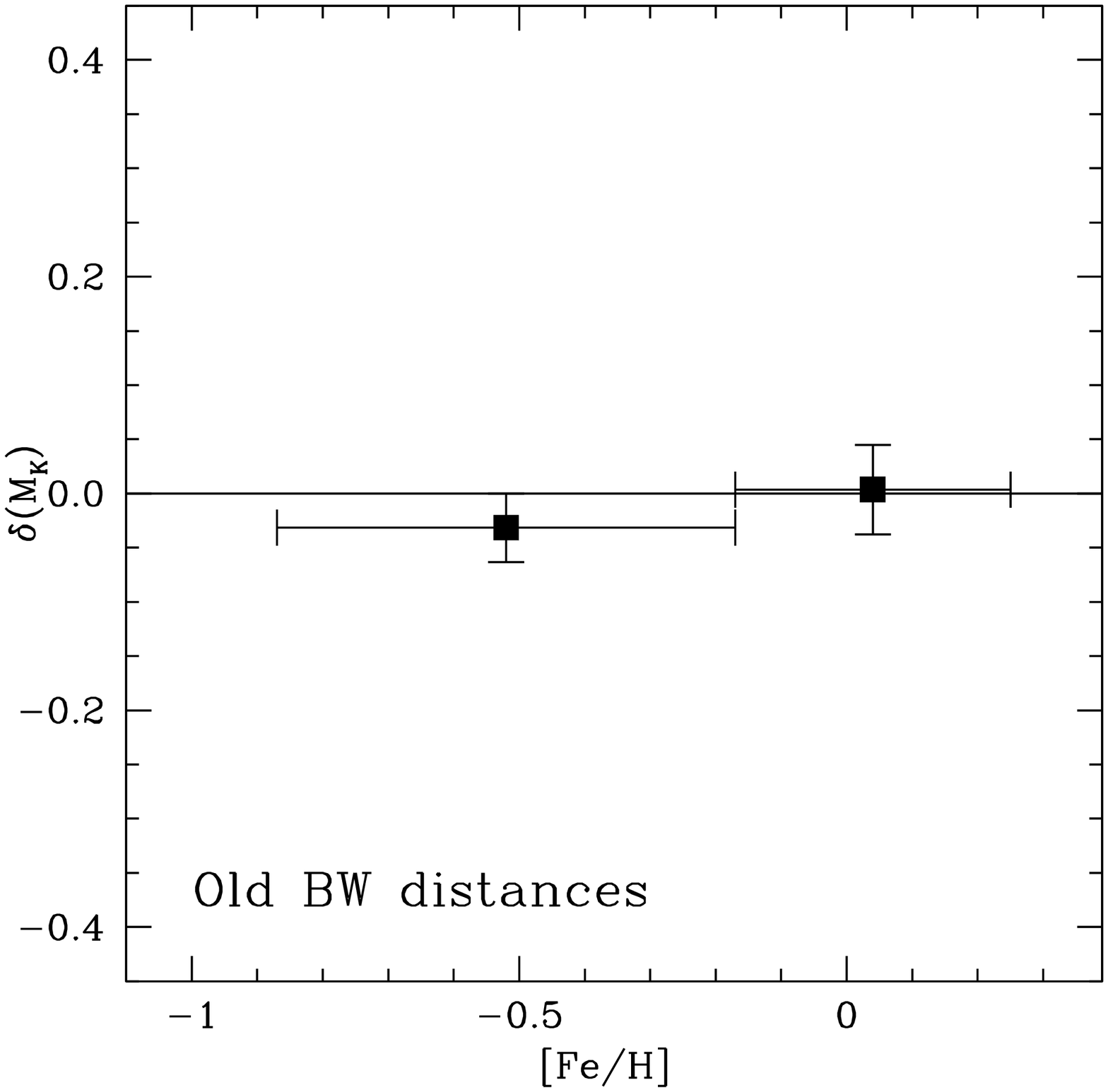}} &
      \resizebox{80mm}{!}{\includegraphics{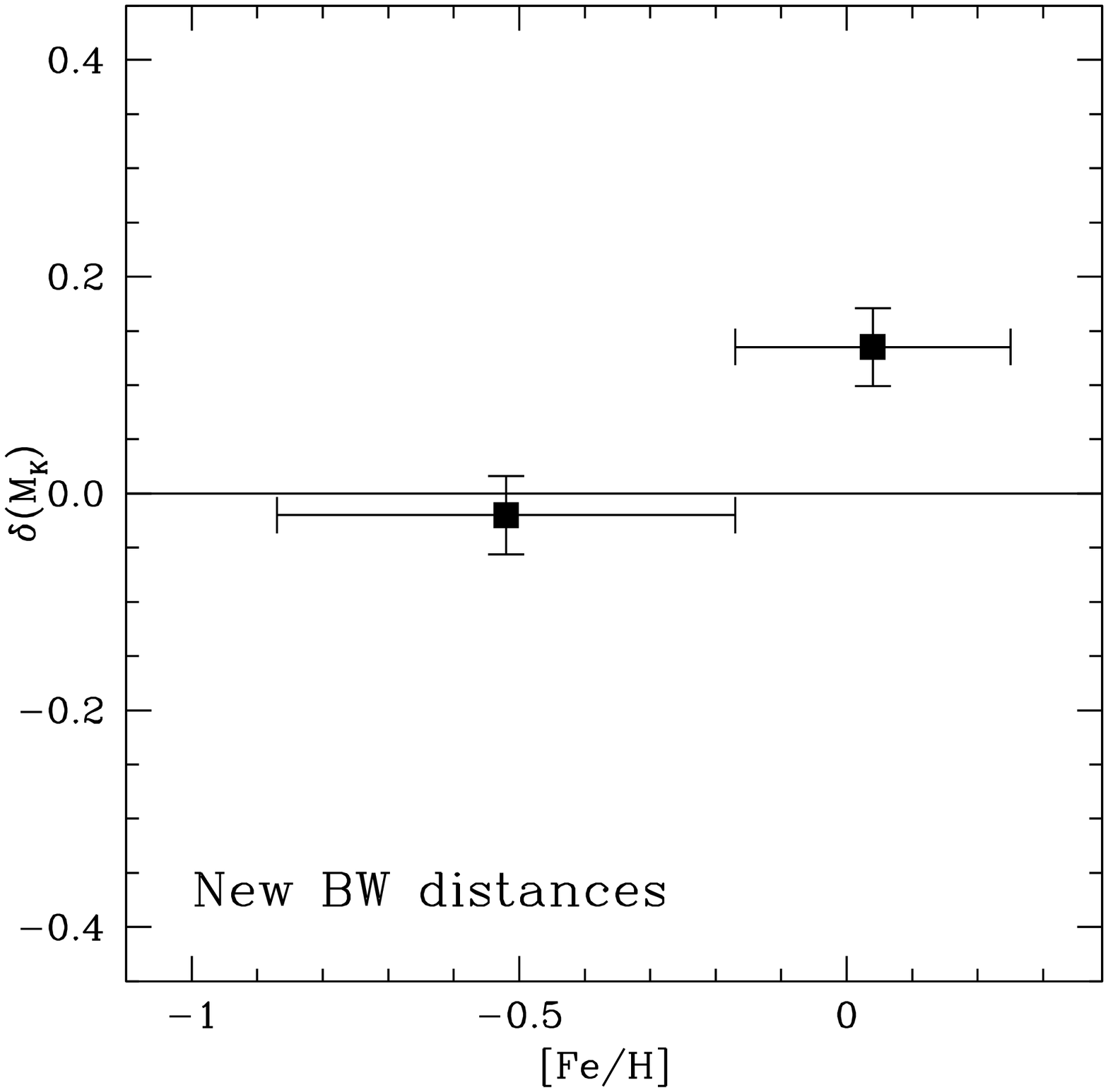}} \\
   \end{tabular}
      \caption{The top panels show the PL relations calculated in each bin for 
the $K$ band. The bottom panels show the residuals $\delta(M_V)$ as a function 
of the iron abundance for both the "Old" and the "New" Galactic Cepheid distances. 
The mean values of $\delta(M_K)$ in each metallicity bin are 
plotted as filled squares, with the vertical error-bars representing its 
associated error. The horizontal bars indicate the dimension of the bins. 
The horizontal solid lines indicate the null value which corresponds to 
an independence of the PL relation from the iron content.} \label{Fig9}
   \end{figure*}

\begin{table*}
\label{table:9}
\caption{Stellar parameters and \ion{Fe}{i}, \ion{Fe}{ii} abundances of our Magellanic 
Cepheids. The \ion{Fe}{i} values were adopted as final [Fe/H]. When available the value from 
previous studies is also reported ([Fe/H]$_L$, where L stands for Luck \& Lambert 1992; 
Luck et al. 1998). In the last column are listed the same values (as in column \# 7), 
but rescaled to the iron solar abundance adopted in our work.}
\begin{center}
\renewcommand{\footnoterule}{}
\begin{tabular}{l l c l r c c c} \hline\hline
ID & $T_{\rm eff}$ & $v_{t}$ & $\log g$ & [\ion{Fe}{i}/H] & [\ion{Fe}{ii}/H]& [Fe/H]$_{L}$ & [Fe/H]$_{L_C}$   \\ \hline
\multicolumn{8}{c}{LMC} \\  \hline
HV 877    &   4690  &  5.40   & 0.5  &  -0.44  &  -0.47  &  \ldots	    &	\ldots       \\
HV 879    &   5630  &  3.05   & 1.0  &  -0.14  &  -0.14  &  -0.56$^{b}$ &	-0.46  \\ 
HV 971    &   5930  &  2.30   & 1.4  &  -0.29  &  -0.29  &  \ldots	    &	\ldots       \\
HV 997    &   5760  &  3.10   & 1.2  &  -0.21  &  -0.22  &  \ldots	    &	\ldots       \\
HV 1013   &   4740  &  5.35   & 0.2  &  -0.59  &  -0.60  &  \ldots	    &	\ldots       \\
HV 1023   &   5830  &  3.10   & 1.1  &  -0.28  &  -0.27  &  \ldots	    &	\ldots       \\
HV 2260   &   5770  &  3.40   & 1.6  &  -0.38  &  -0.36  &  \ldots	    &	\ldots       \\
HV 2294   &   5080  &  3.90   & 0.5  &  -0.42  &  -0.42  &  -0.06$^{a}$ &	+0.10  \\ 
HV 2337   &   5560  &  3.30   & 1.6  &  -0.35  &  -0.36  &  \ldots	    &	\ldots       \\
HV 2352   &   6100  &  3.65   & 1.6  &  -0.49  &  -0.47  &  \ldots	    &	\ldots       \\
HV 2369   &   4750  &  6.00   & 0.3  &  -0.62  &  -0.63  &  \ldots	    &	\ldots       \\
HV 2405   &   6170  &  4.20   & 2.3  &  -0.27  &  -0.28  &  \ldots	    &	\ldots       \\
HV 2580   &   5360  &  2.75   & 0.7  &  -0.24  &  -0.25  &  \ldots	    &	\ldots       \\
HV 2733   &   5470  &  2.90   & 1.8  &  -0.28  &  -0.27  &  \ldots	    &	\ldots       \\
HV 2793   &   5430  &  2.90   & 0.9  &  -0.10  &  -0.11  &  \ldots	    &	\ldots       \\
HV 2827   &   4790  &  4.00   & 0.0  &  -0.38  &  -0.33  &  -0.24$^{b}$ &	-0.14  \\
HV 2836   &   5450  &  2.85   & 1.0  &  -0.16  &  -0.19  &  \ldots	    &	\ldots       \\
HV 2864   &   5830  &  2.80   & 1.5  &  -0.19  &  -0.20  &  \ldots	    &	\ldots       \\
HV 5497   &   5100  &  3.40   & 0.3  &  -0.25  &  -0.24  &  -0.48$^{b}$ &	-0.38  \\
HV 6093   &   5790  &  4.50   & 1.5  &  -0.60  &  -0.60  &  \ldots	    &	\ldots       \\
HV 12452  &   5460  &  2.90   & 1.0  &  -0.35  &  -0.37  &  \ldots	    &	\ldots       \\
HV 12700  &   5420  &  3.15   & 1.4  &  -0.36  &  -0.35  &  \ldots	    &	\ldots       \\ \hline
\multicolumn{8}{c}{SMC}\\ \hline
HV 817    &  5850 &  3.25 &  1.0 &  -0.82 &  -0.84 &  \ldots       &\ldots  	 \\
HV 823    &  5990 &  3.80 &  1.4 &  -0.80 &  -0.81 &  \ldots      & \ldots 	 \\
HV 824    &  5170 &  3.00 &  0.7 &  -0.73 &  -0.74 &  -0.94$^{b}$ &  -0.84   \\
HV 829    &  5060 &  3.30 &  0.2 &  -0.76 &  -0.73 &  -0.61$^{b}$ &  -0.51   \\
HV 834    &  5750 &  2.95 &  1.2 &  -0.63 &  -0.64 &  -0.59$^{b}$ &  -0.49   \\
HV 837    &  5140 &  2.90 &  0.0 &  -0.83 &  -0.80 &  -0.91$^{b}$ &  -0.81   \\
HV 847    &  4790 &  2.80 &  0.0 &  -0.75 &  -0.77 &  \ldots       & \ldots 	 \\
HV 865    &  6130 &  1.90 &  0.5 &  -0.87 &  -0.82 &  -0.44$^{a}$ &  -0.28	 \\
HV 1365   &  5340 &  2.48 &  0.6 &  -0.82 &  -0.84 &  \ldots      & \ldots          \\
HV 1954   &  5890 &  2.47 &  1.0 &  -0.76 &  -0.75 &  \ldots     & \ldots 	 \\
HV 2064   &  5550 &  3.10 &  0.7 &  -0.64 &  -0.64 &  \ldots      & \ldots 	 \\
HV 2195   &  5970 &  2.90 &  1.0 &  -0.67 &  -0.68 &  -0.45$^{a}$ &  -0.29	 \\
HV 2209   &  6130 &  2.30 &  1.1 &  -0.65 &  -0.67 &  \ldots      & \ldots 	 \\
HV 11211  &  4830 &  2.60 &  0.0 &  -0.83 &  -0.81 &  \ldots      & \ldots 	 \\ \hline
\hline
\multicolumn{8}{l}{$^{a}$ Luck \& Lambert (1992).}\\    
\multicolumn{8}{l}{$^{b}$ Luck et al. (1998).}\\
\end{tabular}
\end{center}
\end{table*}

%%%%%%%%%%%%%%%%%%%%%%%%%%%%%%%%%%%%%%%%%%%%%%%%%%%%%%%%%%%%%%%%%%%%%%%%%%%%%%%%%%%%%%%%%%%%%%%%%%
\section{The effect of [Fe/H] on the PL relation}

To assess the effect of the iron content on the Cepheid PL relation we select, 
among our sample, only the stars with periods between 3 and 70 days 
(61 stars out of 68), populating, in this way, the linear part of the PL relation, 
i.e. the one useful for distance determinations (e.g. Bono et al. 1999; 
Marconi et al. 2005). 

By using the values given in the current literature (e.g. 
Benedict et al. 2002; Walker 2003; Borissova et al. 2004; 
Sollima et al. 2008; Catelan \& Cortes 2008; Groenewegen et al. 2008),
we adopted for the barycentre of the LMC a true distance modulus 
($\mu_{LMC}$) of 18.50 mag (with an error of $\pm$ 0.10). This is 
also consistent with the standard PL relations (Freedman et al. 
2001; Persson et al. 2004) used as comparison. The SMC is 
considered 0.44 $\pm$ 0.10 mag more distant than the LMC (e.g. Cioni 
et al. 2000; Bono et al. 2008). This value of the relative distance 
between the two galaxies has confirmed the results of previous studies 
(Westerlund 1997, and references therein). For all the Galactic Cepheids 
we have individual distances, as mentioned in Section 2 above.

It is necessary to divide our Cepheid sample into bins of metallicity to investigate 
its effect on the PL relation. The number of bins needs to be chosen taking into account
two competing effects. On the one hand, dividing the stars in more bins in principle allows
to disentangle finer details. On the other, however, each bin needs to contain enough stars
so that the instability strip is well populated and, hence, spurious sampling effects are negligible.
We choose two bins with about 30 stars each as best compromise between a detailed 
investigation and statistical significance.This choice is justified by the simulations described
in the next section.

Our results for the $V$ and $K$ bands are presented in Figures 8 and 9, 
respectively. In the top panels are shown the PL relations in each metallicity
bin calculated with a linear regression. In each panel  are also indicated the
average iron content of the bin, the root mean  square of the linear regression
and the number of stars.

The bottom panels of Figures 8 and 9 show the magnitude
residuals $\delta(M)$ in the $V$ and $K$ band, respectively, 
as a function of [Fe/H]. These magnitude residuals are calculated 
as the  difference between the observed absolute magnitude and 
the absolute magnitude as determined from its period using a 
standard PL relation, namely the $PL_V$  from Freedman et al. (2001) 
and the the $PL_K$ band Persson et al.  (2004). In practice, 
$\delta(M)$ is the correction to be applied to a
universal, [Fe/H]-independent PL relation to take into account the effects of metallicity.
A positive $\delta$ means that the actual luminosity of a Cepheid is fainter than the one
obtained with the standard PL relation.
The filled squares in the bottom panels of Figures 8 and 9 represent the mean values of
$\delta(M)$ in each metallicity bin, the vertical error-bars are the  errors on the mean
and the solid line is the null value, which corresponds to an independence of the PL
relation on the iron content. The horizontal bars indicate the size of the metallicity bins.

Data plotted in the bottom right panel of Fig.~8 indicate that the magnitude residuals of the 
$V$ band in the individual bins are positive and located at $\approx 2 \sigma$ (metal-poor) 
and $\approx 9 \sigma$ (metal-rich) from zero. Moreover, the $\delta(M_V)$ in the two bins 
differ from each other at the $3\sigma$ level. Our data, then, suggest that metal-rich 
Cepheids in the $V$ band are, at a fixed period, fainter than metal-poor ones.
Data plotted in the bottom left panel indicate that this finding is marginally 
affected by uncertainties in the Galactic Cepheid distance scale. The use of 
Cepheid distances provided by Groenewegen (2008) provides a very similar trend  
concerning the metallicity effect. Moreover, the magnitude residuals based on 
the "Old Sample" are located $\approx 1.5 \sigma$ (metal-poor) and $\approx 4 \sigma$ 
(metal-rich) from zero and the difference in the two bins is larger than one $\sigma$. 

\begin{figure}
   \centering
   \includegraphics[width=8cm]{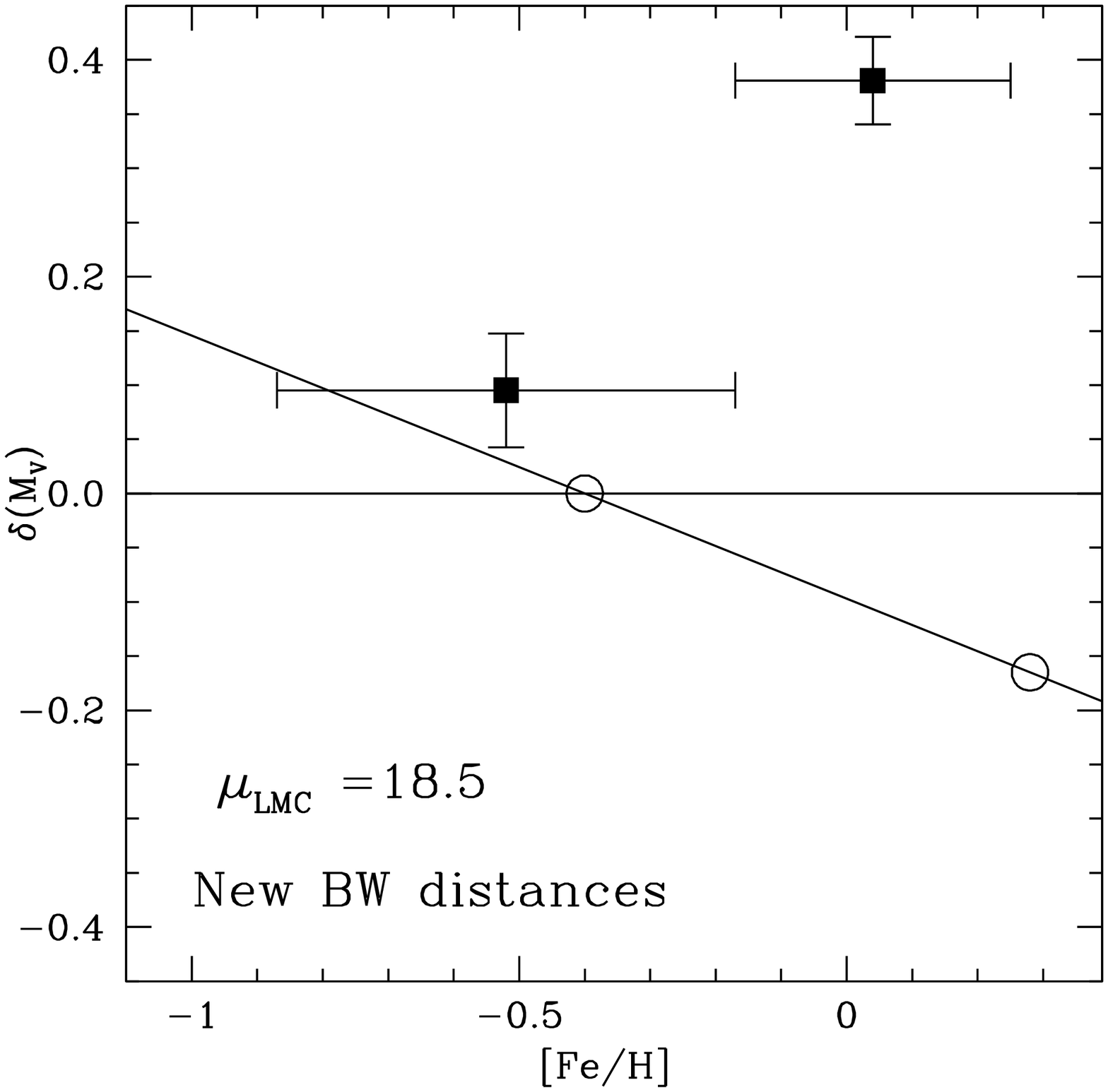}
      \caption{The $V$-band residuals compared to Freedman et al. (2001) PL 
relation are plotted against the iron content measured from observed 
spectra (bottom right panel in Fig.~8). The filled squares display the 
mean value in each metallicity bin, with its associated errorbar. 
The metallicity dependence estimated by Kennicutt et al. (1998) 
using two Cepheid fields in M101 (open circles) is shown as a full line.}
         \label{Fig10}
   \end{figure}

The results for the $K$ band are displayed in Fig.~9. 
The near-infrared data are on different photometric systems, therefore, we 
transformed them into the 2MASS (Two Micron All Sky Survey) photometric 
system. In particular, the apparent $K$ magnitudes listed in Table~2 and 3 
were transformed using the transformations provided by Koen et al. (2007), while 
the $PL_K$ relation by Persson was transformed using the transformations provided 
by Carpenter (2001). The correction are typically of the order of $\sim$ 0.02 mag. 
Data plotted in the bottom right panel of Fig.~9 indicate that the metal-poor bin 
is within $1\sigma$ consistent with zero. On the other hand, the metal-rich bin 
differs from the null hypothesis by $\approx 4\sigma$. Moreover, the magnitude 
residuals in the two metallicity bins differ by $\approx 2 \sigma$. This 
finding is at odds with current empirical (Persson et al. 2004; 
Fouqu\'e et al. 2007) and theoretical (Bono et al. 1999,2008; Marconi et al. 2005) 
evidence suggesting that the $PL_K$ relation is marginally affected by 
metal abundance. In order to constrain this effect we also adopted the 
"Old Sample" distances. The magnitude residuals plotted in the bottom 
right panel of Fig.~9 show that the two metallicity bins are, within 
the errors, consistent with no metallicity effect. Unfortunately, 
current error budget (distance, metal abundance) does not allow 
us to reach firm conclusions concerning the metallicity dependence 
of the $PL_K$ relation.

%%%%%%%%%%%%%%%%%%%%%%%%%%%%%%%%%%%%%%%%%%%%%%%%%%%%%%%%%%%%%%%%%%%%%%%%%%%%%%
\subsection{Estimation of the errors caused by sampling of the PL}
The error on the residuals showed  in Fig.~8 and Fig.~9 is characteristic of
the particular sample of stars we have used. In this section we describe the
simulations we have performed in order to assess the impact of different samplings
of the PL relation.

In order to do so, we have extracted random subsamples composed by different numbers
of stars from a sample of observed Cepheid that  populate well the PL relation. We
choose as reference a sample of 771 LMC  Cepheids from the OGLE database
(Udalski et al. 1999) for the test in the  $V$-band and the sample of 92 LMC Cepheids
(Persson et al. 2004) for the  test in the $K$-band. The latter is the largest 
observed sample of
Cepheids  in the near-infrared bands. For each extraction a linear regression was
performed to derive the slope and the zero point of the resulting PL relation. Also,
we calculated the magnitude residuals $\delta(M_V)$ and $\delta(M_K)$ as defined
above for the actual observed programme stars. We have, then, compared these
quantities with those derived for the whole sample in order to estimate the  error
due to random sampling and to optimize the number of bins we can divide our sample 
into. It is worth noticing here that this procedure will somewhat
\mbox{\it{overestimate}} the sampling error. This is because some of the random
extractions will generate  a subsample that covers a small range of periods, while
we have carefully selected our sample in order to cover a broad period range.

After performing 1 million extractions in each of the bands, $V$ and $K$, for several
bin sizes, we have settled for two metallicity bins of about 30 stars each. With this
choice, $\delta(M_V)$ results to being less that 0.1 mag in 95\% of the cases and never
larger than 0.15. The mean value of the distribution is 0.03 mag. These results imply
that the non-zero result for the high metallicity bin in Figure~\ref{Fig8} cannot be due
to insufficient sampling of the PL relation.

As for the $K$ band, the simulations indicate that $\delta(M_K)$ is smaller that 0.04
mag in 95\% of the cases and never larger than 0.06 mag, with a mean of 0.01 mag.
Also in this case, then, the results discussed above and displayed in Figure~\ref{Fig9}
are not significantly affected by sampling errors.

\subsection{Comparison with previous results}
We compare our results with two different behaviours (see Fig.~10) 
as examples of the effects of the metallicity on the PL relation currently 
available in the literature: independence from the iron content and a 
monotonic decreasing trend (e.g. Kennicutt et al. 1998; Sakai et al. 2004; 
Storm et al. 2004; Groenewegen et al. 2004; Macri et al. 2006; Sandage 
\& Tammann 2008), in the sense that metal-rich Cepheids are brighter 
than metal-poor ones.

We use for the comparison the classical results of Kennicutt et al. 
(1998) adopted by the HST Key Project to determine extragalactic 
distances with Cepheids (Freedman et al. 2001). They have analysed 
two Cepheid fields in M101, with average values of metallicity around 
-0.4 dex and 0.28 dex (determined from measurements of oxygen in H II 
regions in the two fields). They have observed 29 Cepheids in the outer 
field (low metallicity) and 61 Cepheids in the inner field (high metallicity) 
with periods between 10 and 60 days. Considering the outcome of our simulations, 
the two Cepheid samples observed by Kennicutt and collaborators are comparable 
to our sub-samples in each bin, it is then reasonable to compare the qualitative 
indications about the effect of the metallicity that can be derived from the two 
analyses. The complete comparison is possible only in the $V$-band. 
%since there are no data available for the monotonically decreasing $\delta$(M) 
%hypothesis in the $K$-band. 

As we have already mentioned the error due to the sampling of the PL relation 
is much smaller than the one on the residuals. Data plotted in Figures 8,9 
and 10 disclose several circumstantial evidence.  
This means that the increasing trend of the latter in the $V$-band, as a function of the 
iron content, is real with a confidence level of 99\%.\\

{\em i)} $V$-band case

\begin{itemize}
\item {\em No dependence of $\delta(M_V)$ on [Fe/H]}: 
A null effect on the metal abundance would imply that the residuals of 
the two bins should be located, within the errors, along the independence line.
Current findings and the outcome of the simulations mentioned above indicate that 
this hypothesis can be excluded completely. The increasing trend of the residual 
in the $V$-band, as a function of the iron content, is real with a confidence 
level larger than 95\%. \\
\\

\item {\em Monotonically decreasing $\delta(M_V)$}: we compare the classical 
results of Kennicutt et al. (1998, open circles and solid line in Fig.~10) 
with our data.  Since the increasing trend of our residuals is real, this 
hypothesis is incompatible with our results. 
In passing, we also note that the reddening estimates adopted by 
Kennicutt et al. (1998) are based on the observed mean colors. However, 
metal-poor Cepheids are, at fixed period, hotter than metal-rich ones. 
Therefore, Kennicutt et al. would have underestimated the reddenings and
the luminosities of his metal-poor sample, producing an apparent
underluminosity for metal-poor Cepheids.\\
\\

\end{itemize}

{\em ii)} $K$-band case

\begin{itemize}
\item Current data do not allow us to reach a firm conclusion concerning the 
metallicity effect.\\
\end{itemize}

To summarize, we found an increasing trend of the $V$-band residuals with 
the iron content.  This result is in disagreement with an independence of 
the PL relation on iron abundance and with the linearly decreasing trend 
found by other observational studies in the literature (e.g. 
Kennicutt et al. 1998).

%%%%%%%%%%%%%%%%%%%%%%%%%%%%%%%%%%%%%%%%%%%%%%%%%%%%%%%%%%%%%%%%%%%%%%%%%%%%%%%
\subsection{LMC distance: "short" scale vs "long" scale}

Regrettably we can find in the recent literature values 
of the LMC distance modulus ranging from 18.1 to 18.8, not 
always obtained with different techniques. Those studies 
finding a distance modulus less than 18.5 support the so-called 
"short" distance scale, whereas those finding it greater than 
18.5 support the "long" distance scale (for a review of the 
results and methods see Benedict et al. 2002; Gibson 2000). 
More recently, Schaefer (2008) found that distance estimates to 
the LMC published before 2001 present a large spread 
( $ 18.1 \le \mu \le 18.8$). On the other hand, distances published 
after 2002 tightly concentrate around the value adopted by the HST 
Key Projects ($\mu =18.5$, Freedman et al. 2001; Saha et al. 2006).      
In order to overcome this suspicious bias, we decided to investigate 
the different behaviour of the PL relation depending on the adopted 
LMC distance scale.
In order to do that, we have repeated our calculation of the $V$-band 
residuals assuming a distance modulus for LMC of 18.3 (representative 
of the "short" scale) and of 18.7 (for the "long" scale). The results 
are shown in Fig.~11.

\begin{figure*}
  \begin{center}
    \begin{tabular}{cc}
      \resizebox{80mm}{!}{\includegraphics{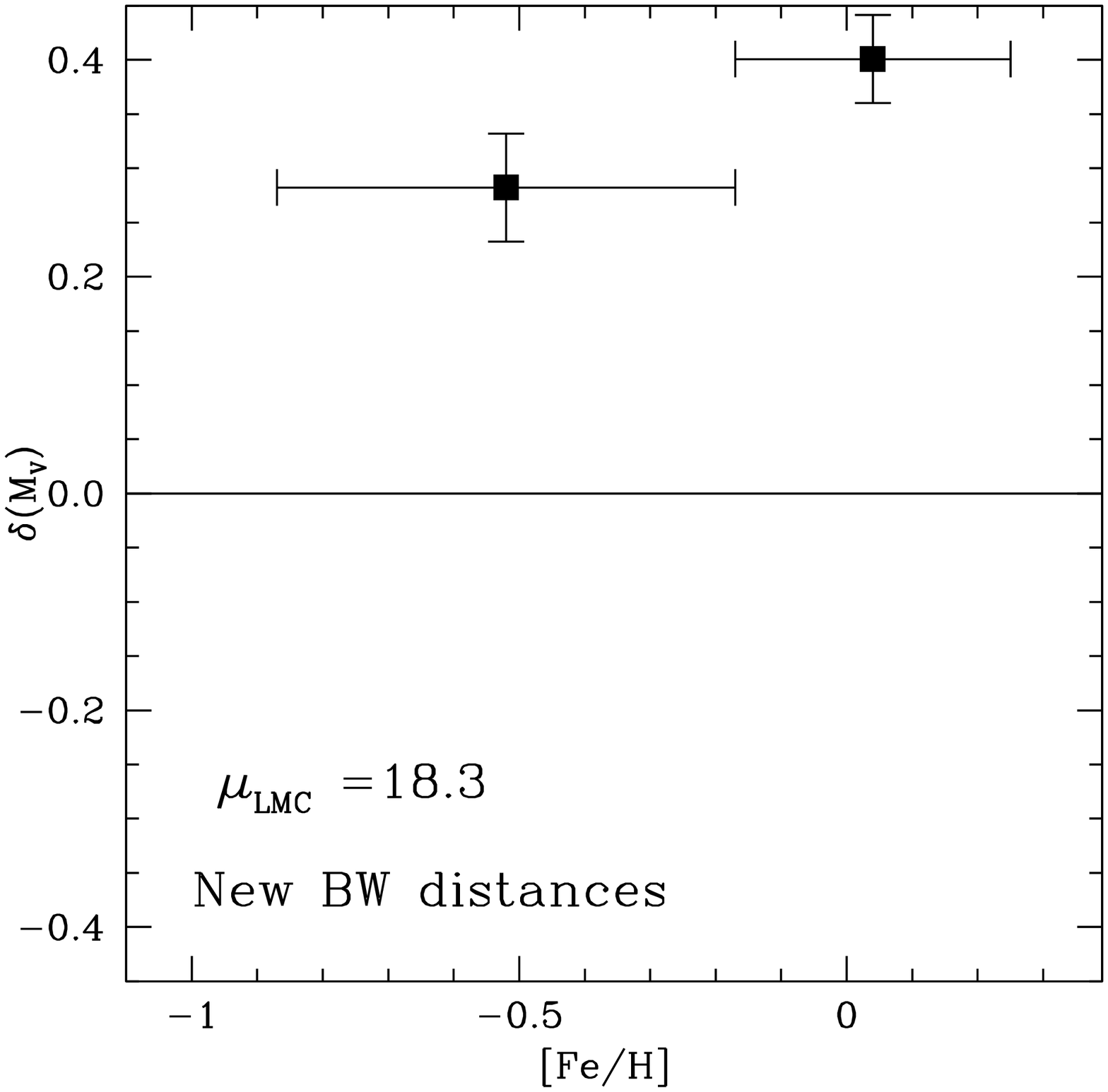}} &
      \resizebox{80mm}{!}{\includegraphics{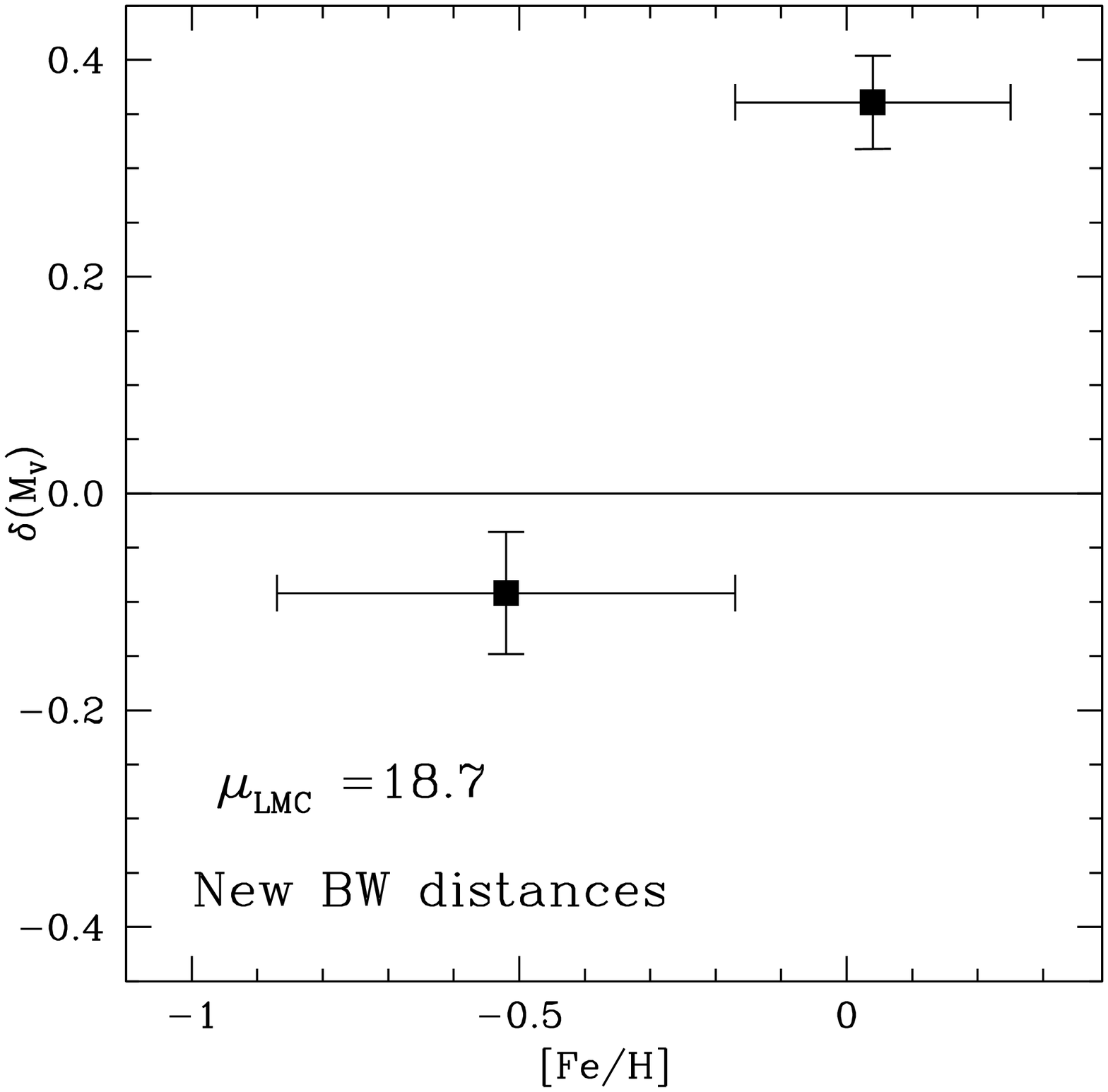}} \\
    \end{tabular}
   \end{center}
   \caption{The $V$-band residuals compared to the Freedman et al. (2001) 
PL relation versus the iron content measured from observed spectra, assuming 
an LMC distance modulus of 18.3 (left) and of 18.7 (right). The filled squares 
represent the mean value in each metallicity bin, with its associated errorbar.} 
   \label{Fig11}
\end{figure*}
In the left and right panel, the $V$-band residuals refer to the 
"short" and the "long" distance scale, respectively. 
For $\mu_{LMC}$ of 18.3, the $V$-band residuals are located at 
least at 5 $\sigma$ from zero. The difference between the 
metal-poor and the metal-rich bin is of the order of one $\sigma$. 
This trend is similar to the trend we obtained using $\mu_{LMC}$=18.5, 
however, the distance from zero of the metal-poor bin is significantly 
larger. This result disagrees with an independence of the 
PL relation on the iron abundance and the monotonic decreasing behaviour 
at the 99.99 \% level (according to the $\chi^2$ method). On the other hand,
the $\delta(M_V)$ values for $\mu_{LMC}$ of 18.7 are located at $\approx 2$ 
(metal-poor) and $\sim$ 8 (metal-rich) $\sigma$ from zero. The difference 
between the two bins is at least at 4 $\sigma$ level. The data trend is 
slightly steeper than for $\mu_{LMC}$=18.5. Using again a $\chi^2$ technique, 
we find that this result disagrees with an independence of the 
PL relation on iron abundance and with the linearly decreasing trend 
often quoted in the literature (e.g. Kennicutt et al. 1998).

The results based on the tests performed assuming different LMC distances 
are the following:\\
{\bf Short scale -} Data plotted in the left panel of Fig.~11 indicate
that $V$-band PL relation does depend on the metal content. Indeed, the 
two bins are located at least at 5 $\sigma$ from zero. This means that 
the zero-point of the quoted PL relations do depend on the metal 
abundance. However, the difference between the two bins is small 
and of the order of one $\sigma$.  This indicates that the metallicity 
effect supported by the short scale is mainly caused by a difference 
in the zero-point. This result would imply a significant difference 
in the zero-point of Magellanic Cepheids. However, such a difference 
is not supported by current empirical (Laney \& Stobie 2004; 
van Leeuwen et al. 2007; Fouqu\'e et al. 2007; Sandage \& Tammann 2008) 
and theoretical evidence. 

{\bf Long scale -} Data plotted in the right panel of Fig.~11 indicates 
that $V$-band PL relation does depend on the metal content. 
The difference from zero ranges from $\sim 2$ for the metal-poor 
bin to more than 8 $\sigma$ for the metal-rich bin. Moreover, the 
two bins differ at $4.5 \sigma$ level. This means that the metallicity 
effect supported by the long scale is mainly caused by a difference 
in the slope. 
This finding, taken at face value, would imply that metal-poor Cepheids 
(e.g. Cepheids in IC~1613, $-1.3 \le [Fe/H]\le -0.7$, Skillman et al. 2003) 
should be, at fixed period, $\approx 0.5$ ($V$) mag brighter than Galactic 
Cepheids. This difference is not supported by current empirical (Dolphin 
et al. 2003; Antonello et al. 2006; Pietrzynski et al. 2006; 
Saha et al. 2006; Fouqu\'e et al. 2007) and theoretical evidence. 

The quoted results suggest that the $V$-band PL relation is affected by 
metal abundance. This finding is marginally affected by the adopted LMC 
distance. It is worth mentioning that recent empirical estimates based 
on robust primary standard candles indicate that the true LMC distance 
is $18.5\pm 0.1$ (Alves 2004; Benedict el al.  2007; van Leeuwen et al. 2007; 
Catelan \& Cortes 2008; Feast et al. 2008; Groenewegen et al. 2008; 
Sollima et al. 2008). In view of this convergence on the LMC distance 
and on the results based on the "short" and on the "long" distance scale, 
the results based on $\mu_{LMC}=18.5$ appear to be the most reliable ones.

%%%%%%%%%%%%%%%%%%%%%%%%%%%%%%%%%%%%%%%%%%%%%%%%%%%%%%%%%%%%%%%%%%%%%%%%%%%%
\section{Conclusions}

We have directly measured the iron abundances for 68 Galactic and Magellanic 
Cepheids from FEROS and UVES high resolution and high signal-to-noise spectra. 
We have used these measurements to assess the influence of the stellar iron 
content on the Cepheid PL relation in the $V$ and in the $K$ band. 
In order to do this we have related the $V$-band and the $K$-band residuals 
from the standard PL relations of Freedman et al. (2001) and 
Persson et al. (2004), respectively, to [Fe/H].
Differently from previous studies, we can constrain the PL relation using 
Cepheids with known distance moduli and chemical abundances, homogeneously 
measured, that cover almost a factor of ten in metallicity.

For our Galactic sample, we find that the mean value of the iron content 
is solar ($\sigma$ = 0.10, see Fig.~6), with a range of values between -0.18 dex and
+0.25 dex. For the LMC sample, we find that the mean value is about $\sim$
-0.33 dex ($\sigma$ =0.13, see Fig.~6), with a range of values between -0.62 dex and
-0.10 dex. For the SMC sample, we find that the mean value is about $\sim$ 
-0.75 dex ($\sigma$ = 0.08, see Fig.~6), with a range of values between -0.87 
and -0.63.

We have compared our results with the analyses of FC97, Andrievsky et al. 
(2002a, 2002b, 2002c) and Luck et al. (2003) for the Galactic Cepheids 
and LL92 and L98 for the Magellanic Clouds. Regarding the Galactic sample, 
our results are marginally more in agreement with Andrievsky's values than 
with FC97 and the differences, on average, appear rather small. 
Considering the Magellanic Cepheids, we have a poor agreement with LL92, 
which could be in part accounted for by different analytical tools 
and data quality. Our data are in better agreement with L98 results and 
the spread of our iron abundances in the LMC and SMC is similar to the one 
they reported. We note that the mean metallicity that they found with their 
complete sample (-0.30 dex and -0.74 dex) is in very good agreement with our 
results.

Our main results concerning the effect of the iron abundance on the PL relation 
are summarized in Fig.~8 and Fig.~9 (bottom panels) and they hold for a 
LMC distance modulus of 18.50. In Fig.~10 is also showed the comparison in the 
$V$-band with the empirical results of Kennicutt et al. (1998) in two 
Cepheid fields of M101 (open circles and solid line). The main findings can 
be summarized as follows:

\begin{itemize}
\item The $V$-band PL relation does depend on the metal abundance. 
This finding is marginally affected by the adopted distance scale for 
the Galactic Cepheids and by the LMC distance. 
 
\item Current data do not allow us to reach a firm conclusion concerning 
the dependence of the $K$-band PL relation on the metal content. The 
use of the most recent distances for Galactic Cepheids (Benedict et al. 2007; 
Fouqu\`e et al. 2007; van Leeuwen et al. 2007) indicates a mild metallicity 
effect. On the other hand, the use of the old distances (Storm et al. 2004) 
suggest a vanishing effect. 

\item Residuals based on a canonical LMC distance ($\mu_{LMC}=18.5$) 
and on the most recent distances for Galactic Cepheids present a 
well defined effect in the $V$-band. The metal-poor and the metal-rich bin 
are $\approx 2 \sigma$ and $\approx 9 \sigma$ from the null hypothesis.
Moreover, the two metallicity bins differ at the $3\sigma$ level. 

\item By assuming a "short" LMC distance ($\mu_{LMC}=18.3$) the residuals 
present a strong metallicity dependency in the zero-point of the $V$-band 
PL relation. 
By assuming a "long" LMC distance ($\mu_{LMC}=18.7$) we found a strong 
metallicity effect when moving from metal-poor to metal-rich Cepheids. This 
indicates a significant change in the slope and probably in the zero-point. 
The findings based on the "short" and on the "long" LMC distance are at odds 
with current empirical and theoretical evidence, suggesting a smaller 
metallicity effect. 

\item Metal-rich Cepheids in the $V$-band are systematically fainter than 
metal-poor ones. This evidence is strongly supported by the canonical, 
the "short" and the "long" LMC distance. 

\end{itemize}

The above results together with recent robust LMC distance estimates 
indicate that the behaviours based on the canonical distance appear
to be the most reliable ones. 

In order to constrain on a more quantitative basis the metallicity 
dependence of both zero-point and slope of the optical PL relations  
is required a larger number of Cepheids covering a broader range in 
metal abundances. Moreover, for each metallicity bin Cepheids 
covering a broad period range are required to reduce the error on 
the residuals and to constrain on a quantitative basis the fine 
structure of the PL relation in optical and NIR photometric bands.

\begin{acknowledgements}
It is a pleasure to thank R.E. Luck for sending us his calibrated 
spectra in electronic form.
We warmly thank E. Pompei for carrying out part of the 
FEROS observations for us and J. Storm for many useful insights 
on recent Cepheid distance determinations. Special thanks also 
go to A. Weiss for many useful discussions and suggestions along 
the entire project. We gratefully acknowledge an anonymous referee 
for his/her constructive remarks which helped to improve the content 
and the readability of the paper. We also thank the A\&A editorial 
office for his support. One of us (GB) acknowledges support from the 
ESO Visitor program. This project was partially supported by ASI
(P.I.: F. Ferraro) and by INAF (P.I.: M. Bellazzini).  
This paper has had a complicated story that can be perfectly 
summarized by a latin sentence from Cicero ({\em De Inventione, 
Liber Secundus, 163--164}) 
\begin{quote} 
{\tt Patientia est honestatis aut utilitatis causa rerum arduarum 
ac difficilium voluntaria ac diuturna perpessio; perseverantia est 
in ratione bene considerata stabilis et perpetua permansio.}    
\end{quote} 

\end{acknowledgements}

\end{document}